\documentclass[twocolumn,showpacs,preprintnumbers,amsmath,amssymb]{revtex4}

\usepackage[unicode]{hyperref}
\usepackage[caption=false]{subfig}%
\usepackage{epsfig}
\usepackage{graphicx}
\graphicspath{ {./images/} }
\usepackage{dcolumn}
\usepackage{bm}

\begin{document}

\title{Similarity and self-similarity in random walk with fixed, random and shrinking steps}
\author{ Tushar Mitra$^1$, Tomal Hossain$^1$, Santo Banerjee$^2$ and Md. Kamrul Hassan$^1$}
\affiliation{
$1$ University of Dhaka, Department of Physics, Theoretical Physics Division, Dhaka 1000, Bangladesh \\
$2$ Department of Mathematical Sciences, Giuseppe Luigi Lagrange, Politecnico di Torino, Corso Duca degli Abruzzi 24, Torino, Italy
}

\begin{abstract}

In this article, we first give a comprehensive description of random walk (RW) problem focusing 
on self-similarity, dynamic scaling and its connection to diffusion phenomena. 
One of the main goals of our work is to check how robust  the RW problem is under various 
different choices of the step size. We show that RW with random step size or 
uniformly shrinking step size is exactly the same as for RW with fixed step size. Krapivsky and
Redner in 2004 showed that RW with geometric shrinking step
size, such that the size of the $n$th step is given by $S_n=\lambda^n$ with  a 
fixed $\lambda<1$ value, exhibits some interesting features which are different from the RW 
with fixed step size. Motivated by this, we investigate what if $\lambda$ is not a fixed number 
rather it depends on the step number $n$? To this end, we first generate $N$ random numbers 
for RW of $t=N$ which are then arranged in a descending order so that the size of the $n$th 
step is  $\lambda_n^n$. We have shown, both numerically and analytically, that 
$\lambda_n=(1-n/N)$, the root mean square displacement increases as $t^{1/4}$ 
which are different from all the known results on RW problems.

\end{abstract}

\pacs{61.43.Hv, 64.60.Ht, 68.03.Fg, 82.70.Dd}


\maketitle

\section{Introduction}

Finding order in the disorder has always been an attractive proposition for physicists. 
To that endeavour, physicists have come up with various elegant ideas like the idea of
similarity, self-similarity, scaling, scale-invariance, renormalization group, etc. 
\cite{ref.barenblatt, ref.hassan_santo, ref.redner_krapivsky}. 
These ideas have always been proved to be extremely useful in gaining deep insights into  
large and complex systems. Many of the physical systems are not static rather they evolve 
either probabilistically or deterministically with time. Often we have to extrapolate properties
of such systems for the infinitely long-time limit by taking data from a few snapshots at 
short or intermediate time-limit. Sometimes, we also need to know results of experiment
or simulation for infinitely large systems. However, in reality we can neither do experiment 
nor simulation
in the computer in such large systems. We can still extrapolate results for infinite systems
from the results of a set of data obtained for finite size systems which is known
as finite-size scaling \cite{ref.Hassan_Rahman_1, ref.hassan_didar, ref.hassan_sabbir}. 
The idea of extrapolation is not new. Even Galileo Galilei did that in his famous 
thought experiment while thinking of the now known Newton's
first law. Think of a block being pushed along a surface of a table and observe the distance travelled. 
Now polish the surface of the block and the
table and then apply the same force again. The object then will travel further than before. If we 
continue to polish the surface of the block
and that of the table more and more the object will travel further and further. 
In the end, it can be concluded that if the friction were totally 
absent the block would continue to move forever and ever preserving the same speed. This is called 
extrapolation. The idea of
similarity, self-similarity, scaling and renormalization group essentially helps us to 
gain the ability or the power to extrapolate. 

In fact, the idea of similarity and self-similarity is the key to understand many natural and man-made 
phenomena. In a way, the word self-similarity needs no explanation. Perhaps the best way to 
perceive the meaning of self-similarity is to consider an example. To that end there can be no better
example than the vegetable cauliflower that we all know. The cauliflower head contains branches or 
parts, which when removed and compared with the whole, after blowing it up to a suitable size, 
are found to be apparently the same. 
These isolated branches can again be decomposed into smaller parts, which again look very similar to the whole 
as well as of the branches. Such self-similarity can easily be carried through for about three to four stages
by hand. May be we can go up to a few more steps if we use a microscope.   
After that the structures are too small to go for any further dissection. 
However, from the mathematical point of view, the property of self-similarity may be continued 
through infinite stages. Self-similarity can also be found in branching patterns of snowflakes
and in aggregating colloidal particles which are examples of 
statistically self-similar fractals \cite{ref.hassan_santo}.  Our bodies, like our kidneys, lungs, and circulatory systems 
have a form of self similarity. In fact, it is so widespread in nature that they are extremely 
easy to find. In fact nature chooses to have self-similar objects through evolution in time. Nature
love simplicity. Simple rules when applied over and over again can emerge as an object which
may appear mighty complex. Yet, they are effectively simple and efficient for the purpose they are grown because of their self-similar nature. 

Random walk problem perhaps is the best known example of self-similarity. It has been found
that the dynamics of the random walk problem is governed by diffusion equation suggesting that it
is an important class of stochastic process. It has far reaching
implications as it provides the connection between the theory of random walk and Brownian motion. Owing to this connection we can simulate the diffusion phenomena in the computer. To this end, the idea of random
walk has been used in physics, computer science, ecology, economics and other seemingly disparate 
fields. Typically, a random walk is a sequence of successive random steps. On the
other hand, the random motion of a heavy particle suspended in a bath of light particles is 
known as Brownian motion. It can be described by Langevin dynamics, which replace the collisions with
the light particles by an average friction force proportional to the velocity and a
randomly fluctuating force with zero mean and infinitely short correlation time. Random walk
was first described in literature when the influential journal {\it Nature} published 
a discussion between Pearson and Rayleigh in 1905 \cite{ref.pearson, ref.rayleigh}. It is this 
discussion that attracted physicists like Einstein and Smoluchowski to the subject who 
later made invaluable contribution to it \cite{ref.einstein, ref.smoluchowski}. Random walk with 
random steps, specially 
cases in which the length of the $n$th step changes systematically with $n$, has been 
extensively studied
for due reason of course. Firstly,  the random walk, where step size shrinks following geometric series,
gives rise to a variety of beautiful and unanticipated features which has pedagogical importance \cite{ref.jessen, ref.kershner, ref.winter, ref.erdos,ref.garsia}. Secondly, there also are a variety of situations
where random walks with variable step size are relevant \cite{ref.barkai, ref.h_weiss}.

In this article, we give a comprehensive account of the theory of self-similarity in the light 
of Buckingham $\Pi$-theorem. In particular, we demonstrate that dynamic scaling, which has always been regarded as an ansatz and used as a litmus test for self-similarity, is deeply rooted to the 
Buckingham $\Pi$-theorem \cite{ref.hassan_dynamic_scaling, ref.hassan_liana, ref.hassan_liana_debashish}. 
It is well-known that the random walk problem with fixed step size, which is 
governed by the famous diffusion equation, exhibits self-similarity. Our primary goal is to verify it 
by extensive Monte Carlo simulation using the idea of data-collapse. We also show that
random walk with random step size follows exactly the same solution which proves how robust
 the random walk is {\it vis-a-vis} diffusion phenomena. We then study the random walk problem
for time $t=N$ with shrinking steps such that the size of the $n$th step is $R_n^n$ where $R_n$ 
is defined as follows. We first draw $N$ number of random number
within $[0,1]$ which we then sort in a descending order such that $R_n$ is the $n$th value.
 Krapivsky and Redner have extensively studied them for a wide range of fixed 
$R$ including the case of golden number $(\sqrt{5}-1)/2$ which gives rise to non-trivial 
results \cite{ref.redner_shrinking}. We find completely new results for random walk with 
shrinking steps as we find that the probability
distribution function still obeys the dynamic scaling but with different exponent. In particular,
we find that the root mean square displacement increases with time as $t^{1/4}$ instead
of $t^{1/2}$ for random or fixed step size.

The paper is organized as follows: in Sec. II, we discuss the idea of self-similarity
in the light of Buckingham $\Pi$ theorem and cite an example to explain it.
In Sec. III, a connection between the dynamic scaling and the Buckingham $\Pi$ theorem is shown.
Sec. IV contains a theoretical approach to show that the random walk problem is actually governed
by diffusion equation. In Sec. V, we present a solution of the diffusion equation following the prescription of Buckingham $\Pi$ theorem. Numerical results are presented to verify the analytical
solution for random walk with various different step size in Sec VI  and random walk with shrinking
step size in VII. The results are discussed and conclusions drawn in VIII.

\section{Self-similarity and Buckingham $\Pi$ theorem} 

In physics, we often investigate physical problems or phenomena where it is not enough 
to rely on our naked eyes to judge whether the system possesses self-similarity or not. In fact, it 
might be not that straightforward either. Besides, scientists in general and physicists in 
particular always look for instrumental or mathematical tools to test 
similarity and self-similarity. Often researchers investigate a given phenomena with the aim of 
finding fundamental rules and laws behind it. Such
laws are nothing but relations between a governed quantity, say $a$ and a set of governing 
parameters, say they are $a_1,a_2, ...,a_n$ upon which $a$ depends. Such relations
can always be represented in the form  
\begin{equation}
\label{eq:1}
a=f( a_1, a_2,...,a_n),
\end{equation}
where the quantities $a_1, a_2,...,a_n$ are called the governing parameters. It is then often 
possible to classify the governing parameters $a_1,...,a_n$ into two groups using the definition 
of dependent and independent variables. To that end, consider that the 
parameters $a_{k+1},..., a_n$ have independent dimensions in the sense that none of these 
parameters can be expressed in terms of the power product of the dimensions of the remaining 
parameters $a_1, a_2,..., a_k$ which we call dependent variables.

It implies we can define a dimensionless variable for each dependent variable
 \begin{equation}
\label{eq:2}
\Pi_i={{a_i}\over{a_{k+1}^{\alpha_i}a_{k+2}^{\beta_i}...a_n^{\gamma_i}}}, 
\end{equation}
where $i=1,2,..,k$.
Similarly, we can 
also define a dimensionless quantity for the governed parameter $a$ as well
\begin{equation}
\label{eq:3}
\Pi={{a}\over{a_{k+1}^{\alpha}a_{k+2}^{\beta}...a_n^{\gamma}}}.
 \end{equation}
It essentially can be written as
\begin{equation}
\label{eq:4}
\Pi=F(a_{k+1},a_{k+2},...,a_n,\Pi_1,...,\Pi_k),
 \end{equation}
where $F$ is defined as another function which depends on $n-k$ number of parameters instead
of $n$ number of parameters. Note that $\Pi$ cannot
depend on dimensional variables like $a_{k+1},a_{k+2},...,a_n$ since according to the
definition of dimensionless quantity its numerical value must remain the same even if we
change any or all of the $a_{k+1},a_{k+2},...,a_n$ variables. It means
\begin{equation}
\label{eq:5}
a=a_{k+1}^{\alpha}...a_n^{\gamma}\phi(\Pi_1,\Pi_2,...,\Pi_k).
 \end{equation}
This is known as the Buckingham $\Pi$-theorem \cite{ref.barenblatt, ref.hassan_santo}. 
To find out how it helps to understand the idea of 
similarity and self-similarity we have to first understand the much known geometric similarity and 
then we can extend it to some physical systems.

\begin{figure}
\centering

\subfloat[]
{
\includegraphics[height=3.5 cm, width=4.0 cm, clip=true]
{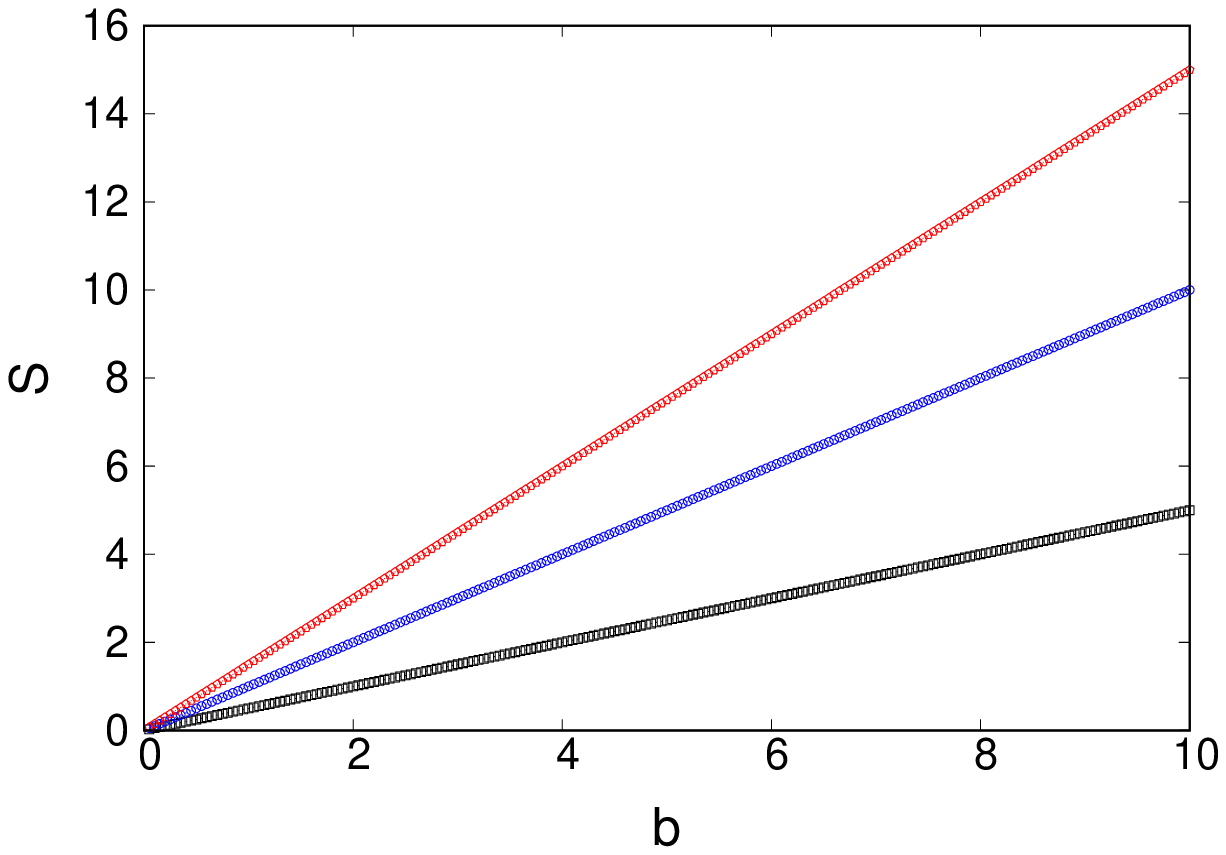}
\label{fig:1a}
}
\subfloat[]
{
\includegraphics[height=3.5 cm, width=4.0 cm, clip=true]
{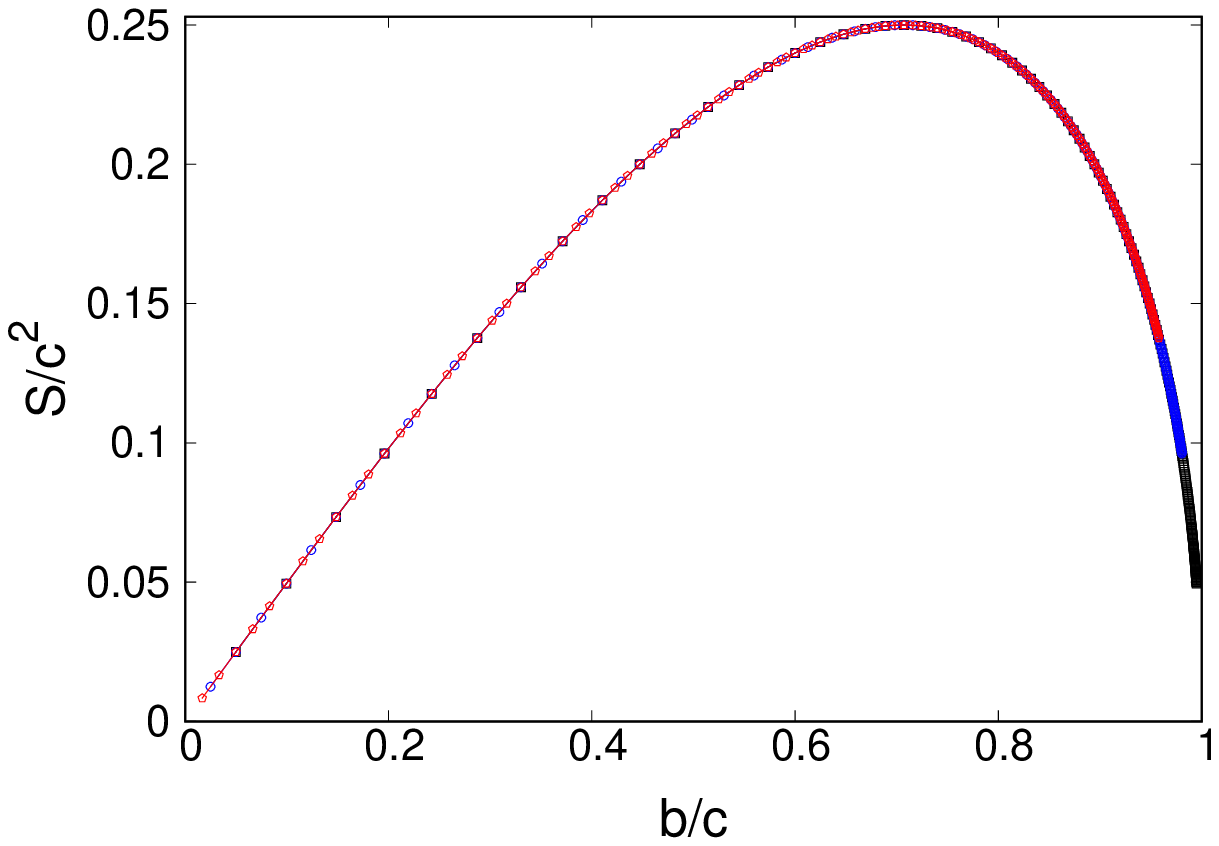}
\label{fig:1b}
}

\caption{(a) This figure shows how area $S$ changes for three different triangles
as a function of height $b$.
 (b) This shows that if we plot $S/c^2$ vs $b/c$ instead of $S$ vs $b$ then all the three curves of (a) 
collapse superbly on to a single master curve. It implies that
for a given numerical value of the ratio $b/c$ the corresponding $S/c^2$ is the same regardless of the
size of the sides of the triangle. 
} 
\label{fig:1ab}
\end{figure}

We all learn the idea of geometric similarity in high school. Let us re-visit 
the same using the formalism of the  Buckingham $\Pi$-theorem. 
Consider that we have a right triangle of sides $a$ (adjacent), $b$ (opposite) and $c$ (hypotenuse). Say, that
we want to measure its area $S$. That is, $S$ is the governed parameter and the sides $a,b, c$ are the governing
parameters. We keep the side $a$ fixed and measure the area $S$ as we vary the side $b$. Obviously,
the size of the hypotenuse $c$ will change as we vary the side $b$ 
and hence in analogy with Eq. (\ref{eq:1}) we can write
\begin{equation}
\label{eq:5}
S=S(b,c).
\end{equation}
A trivially simple dimensional analysis immediately reveals that one of the two governing parameters 
can have independent dimension and let us assume that it is the variable $c$ so that $b$ as well as $S$ 
can be expressed in terms of $c$ alone. We can thus define the following
dimensionless variables.  One, for the dependent variable $b$
\begin{equation}
\label{eq:6}
\xi=b/c, 
\end{equation}
so that the length $b$ is measured in terms of $c$. The other dimensionless
variable is for the governed variable $S$
\begin{equation}
\label{eq:7}
\Pi=S/c^2,
\end{equation}
so that all the areas are measured in unit of $c^2$. Now, the
 numerical value of $\Pi$ remains the same regardless of
the choice of the unit of measurement of length since it is a dimensionless quantity. However, it still depends on $\xi$ and hence in analogy with Eq. (\ref{eq:4})
we can re-write Eq. (\ref{eq:7}) as
\begin{equation}
\label{eq:8}
S=c^2\phi(b/c).
\end{equation}

The question is now: What does it has to do with the idea of similarity and self-similarity? 
Consider that we have
two more triangles differing in size of their sides but share the same acute angle $\theta$. 
Recall that we keep the base or adjacent $a$ fixed and measure area $S$ for each of the three triangles 
as a function of their sides $b$. If we now plot $S$ as a function of their respective $b$ then
we shall have a set of distinct
straight lines, see Fig. (\ref{fig:1a}), one for each different $a$ value with slope equal to $a/2$ since 
we know $S={{a}\over{2}}b$. 
Let us now express $b$ in unit of $c$
and $S$ in unit of $c^2$. It means that if we now plot $S/c^2$ versus $b/c$ 
we find that all the distinct plots of $S$ versus $b$ curve collapse into one universal curve
as shown in Fig. (\ref{fig:1b}). That is, the
numerical value of $S/c^2$ for a fixed value of $b/c$ or acute angle $\theta$, will
be the same no matter how big or how small the triangles are. 
What is the significance of such data-collapse? The numerical value of $S/c^2$ of all right 
triangles having the 
same $b/c$ will coincide. It means all the right triangles
which share the same $b/c$ value are similar. Note that $S/c^2$ and $b/c$ are dimensionless quantity. 
We can extend this idea of geometric similarity to physical phenomena too. We can say two 
or more systems 
or phenomena are similar
if they differ in the numerical values of their dimensional quantities however the numerical 
values of the 
corresponding dimensionless quantities are the same.

\section{Dynamic scaling and Buckingham $\Pi$-theorem}

Many phenomena which physicists often investigate are not static rather evolve probabilistically with time. 
The resulting systems can often be described by kinetic or rate equation approach.
Mathematicians and physicists often look for scaling or self-similar solution to their respective equations
which is actually the solution in the long-time limit. In this limit, the solution usually assumes a simple 
and universal form. In such systems
one often is interested to see if certain observable quantity, say $f(x,t)$, 
exhibits self-similarity or not. To understand what it really means we can apply 
the Buckingam $\Pi$-theorem like we have done for right triangle. Assume that one of the two variables, say
time $t$ for convenience, is the independent variable so that we can express both $x$ and $f$ in terms
of $t$ alone. It means that we have two dimensionless variables $\xi=x/t^z$ and $\phi=f/t^\tau$
where exponents $z$ and $\tau$  are fixed by the dimensional relations $[t^\tau]=[f]$ and $[t^z]=[x]$.
Note that the numerical value of $f/t^\tau$ for a given value of $x/t^z$ will be independent of the
choice of the unit of measurement of time $t$ but $\phi$ may depend on $x/t^z$ and hence according
to Buckingam $\Pi$-theorem we can write the following
\begin{equation}
\label{eq:dynamic_scaling}
f(x,t)\sim t^\tau\phi(x/t^z),
\end{equation}
where $\phi(\xi)$ is known as the scaling function \cite{ref.Hassan_Dongen, ref.family, 
ref.viscek, ref.march, ref.hassan_ba_dc, ref.hassan_debashish}. Self-similarity means that
 snapshots at different times are similar. However, since the same system at different 
times is similar we regard it as the temporal self-similarity \cite{ref.barenblatt}.

\section{Random walk and diffusion equation}

Diffusion is perhaps one of the most ideal examples of natural phenomena that evolves 
probabilistically with time. 
Typically diffusion phenomena are associated with the movement of atoms or molecules within a gas, a 
liquid, or a solid over a more or less long distance. One of the four papers that Einstein 
wrote in 1905,
that raised him to a height that no one has ever reached in the entire history of science, was on the
Brownian motion, which embodies the diffusion process. Since then Brownian motion motion {\it vis-a-vis} the diffusion
process has always been one of the active field of research in almost every branch of science in general and in physics 
in particular. A diffusing particle is subjected to a variety of collisions that we can 
consider random, 
such that each event that occurs between $t$ and $t+\Delta t$ depends only upon the state of the 
system at time $t$ and independent of the state prior to time $t$. This is the property of Markov process.
One can actually consider the Brownian particle as a random walker \cite{ref.reichl,ref.weiss}. Let us consider
that $P(n\Delta,s\tau)$ is the probability
density for the walker to be at position $x=n\Delta$ at time $t=s\tau$ where we assumed that the walk is on a 
one dimensional lattice of lattice constant $\Delta$ and that the time interval between steps is $\tau$.
We can then rewrite the identity
\begin{equation}
    P(x,t_2)=\int P(y,t_1)P(x,t_2|y,t_1)dy,
\end{equation} 
in the following discrete form 
\begin{equation}
\label{eq:markov3}
P(n\Delta,(s+1)\tau)=\sum_{m=-\infty}^\infty P(m\Delta,s\tau)P(n\Delta,(s+1)\tau|m\Delta,s\tau),
\end{equation}
where $P(n\Delta,(s+1)\tau|m\Delta,s\tau)$ is the transition probability to go from site 
$x=m\Delta$ to site 
$x=n\Delta$ in one step \cite{ref.reichl}. This transition probability for the RW therefore is
\begin{equation}
\label{eq:markov4}
P(n\Delta,(s+1)\tau|m\Delta,s\tau)={{1}\over{2}}\delta_{n,m+1}+{{1}\over{2}}\delta_{n,m-1},
\end{equation}
and Eq. (\ref{eq:markov3}) takes the following form
\begin{equation}
\label{eq:markov5}
P(n\Delta,(s+1)\tau)={{1}\over{2}}P((n+1)\Delta,s\tau)+{{1}\over{2}}P((n-1)\Delta,s\tau).
\end{equation}
The two terms on the right account for the increase in $P(n\Delta, \tau)$
because of a hop from $n+1$ to $n$ and hop from $n-1$ to $n$ respectively.

We find it highly instructive to take continuous space-time limit as well. To this end,
we let $x=n\Delta$, $t=s\tau$, and take the limit $\Delta\longrightarrow 0$,
$\tau \longrightarrow 0$ so that $D\equiv {{\Delta^2}\over{2\tau}}$ then we obtain the following 
differential equation for $P(x,t)$
\begin{equation}
\label{eq:diffusion}
{{\partial P(x,t)}\over{\partial t}}=D{{\partial^2 P(x,t)}\over{\partial x^2}},
\end{equation}
which is the well known {\it diffusion equation} for the probability density $P(x,t)$. Einstein gave a heuristic
derivation of the same diffusion equation that describes how the density 
of Brownian particles $P(x,t)$ at point $x$ at time $t$ evolves with time.
It immediately shows that the random walk problem  can also be seen as Brownian particle.
Appreciating it has far reaching consequence. Brownian motion is ubiquitous in nature.  
It is thus possible to look upon the diffusion problem as a random walk executed by the labeled molecule
assuming that successive displacements suffered by the molecule between collisions are
statistically independent. Upon multiplying on both sides of Eq. (\ref{eq:diffusion}) by $x$ and $x^2$ and integrating over the entire range we get
\begin{equation}
\label{eq:rms}
    x_{{\rm rms}}(t)\sim t^{1/2}  \hspace{0.5 cm} {\rm and}  \hspace{0.5 cm} x_{{\rm mean}}(t)=0,
\end{equation}
respectively where
\begin{equation}
x^2_{{\rm rms}}=\int_{-\infty}^\infty x^2P(x,t)dx,
   \end{equation}
   and 
 \begin{equation}
   x_{{\rm mean}}(t)=\int_{-\infty}^\infty xP(x,t)dx,
\end{equation}
assuming that all the walkers start their walk from $x=0$.

\section{Solution to diffusion equation}

The diffusion equation for the probability density function $P(x,t)$ suggests that it is a function 
of two variables only
\begin{equation}
P=P(x,t),
\end{equation}
since time can be re-scaled as $Dt$.
In order to solve the diffusion equation let us first invoke the idea of simple dimensional analysis.
Within the MLT class, their dimensions are
\begin{equation}
[x]=L \hspace{0.4cm} [D]={{L^2}\over{T}} \hspace{0.4cm} [t]=T \hspace{0.4cm} {\rm and} \hspace{0.4cm} [P]=L^{-1}.
\end{equation}
The above dimensional relation implies that the re-scaled time can be chosen to
have independent dimension and define the following dimensionless quantities
\begin{equation}
\xi={{x}\over{\sqrt{Dt}}} \hspace{0.4cm} {\rm and} \hspace{0.4cm} 
\phi(\xi)={{P(x,t)}\over{(Dt)^\theta}},
\end{equation}
where the exponent $\theta=-1/2$ is required by the 
normalization condition $\int_{-\infty}^\infty P(x,t)dx=1$. 
Following the Buckingham $\Pi$-theorem 
we find that the probability density function $P(x,t)$ assumes a simple universal scaling form 
\begin{equation}
\label{eq:ansatz}
P(x,t)\sim {{1}\over{\sqrt{Dt}}}\phi(x/\sqrt{Dt}),
\end{equation}
where $\phi(\xi)$ is the dimensionless scaling function. 
The structure of this scaling form is highly instructive as it greatly simplifies 
further analysis. 

We now substitute Eq. (\ref{eq:ansatz}) in Eq. (\ref{eq:diffusion})  and find that the 
solution of the partial differential equation
reduces to the solution of an ordinary differential equation for the function $\phi(\xi)$ given by 
\begin{equation}
\label{eq:scalingequation}
\Big [{{d^2}\over{d\xi^2}}+{{\xi}\over{2}}{{d}\over{d\xi}}+{{1}\over{2}}\Big]\phi(\xi)=0.
\end{equation}
Solving it subject to the condition $\int_{-\infty}^\infty \phi(\xi)d\xi=1$ we find 
\begin{equation}
\phi(\xi)=A\exp[-\xi^2/4],
\end{equation}
where $A$ is the integration constant fixed by the normalization condition. 
Substituting this into the normalization condition for $\phi$ 
immediately gives 
$A=1/\sqrt{4\pi}$ and therefore
\begin{equation}
\label{eq:solution_difusion}
P(x,t)={{1}\over{\sqrt{4\pi Dt}}}\exp[-{{x^2}\over{4Dt}}].
\end{equation}
We can easily express it as
\begin{equation}
    P(x,t)\sim t^{-1/2}f(x/t^{1/2}),
\end{equation}
where the dynamic scaling function
\begin{equation}
    f(z)={{1}\over{\sqrt{4\pi D}}}\exp[-{{z^2}\over{4D}}],
\end{equation}
and hence Eq. (\ref{eq:solution_difusion}) obeys dynamic scaling.
Furthermore, note that the solution is symmetric about $x=0$ as it satisfies the condition $P(x,t)=P(-x,t)$.
Using this solution Einstein deduced his famous prediction that the
root mean square displacement of Brownian particles is proportional to the square root of time. Besides,
the solution is scale-invariant in the sense that it can be brought to itself under the following similarity
transformation
\begin{equation}
    P\longrightarrow \lambda^{-b/2} P, \hspace{0.25 cm} x\longrightarrow \lambda^{b/2} x, \hspace{0.25 cm}
    t\longrightarrow \lambda^{b}t,
\end{equation}
 since it is a generalized homogeneous function.

\section{Extensive numerical simulation}

The question is: How can we verify the solution, given by Eq. (\ref{eq:solution_difusion}), of the diffusion
equation {\it vis-a-vis} of the random walk problem?  Consider that we ask
$N$ number of walkers to walk starting from the same initial point, which we call origin, 
along the same line with fair coin in their hands. Each walker is asked
to make $n$ steps. The rules of the random walk are as follows. Before attempting to make a step each 
walker flips their coin. Respective walker then make a step to the right of unit step size $\delta=1$ 
if the upper face of the coin appears head and to the left by the same step size if it is tail. 
Owing to the random 
nature of the random walk problem the final position of all the walkers will not be the same. To make
$n$ steps each walker has to flip their coin $n$ times. Say that out of $n$ outcome, 
$n^+$ of them flipped head and $n^-$ of them flipped tail and hence the final position $x_i$ 
of the $i$th walker is obtained by measuring $x_i=n_i^+-n_i^-$. We then create a data by finding 
the fraction of the total $N$ walkers $P(x,n)\Delta x=m/N$ within the position $x$ and $x+\Delta x$ 
where $m$ is 
the number of walkers found within this range. Effectively,  $P(x,n)\Delta x$ represents
the probability that the number of walkers is within the position $x$ and $x+\Delta x$ 
at the end of $n$ steps. 
If we assume each step is made in one unit time then the number of steps $n$ is the time $t$ and 
if we consider 
continuum limit then the solution becomes exactly the same as the solution of the diffusion equation.

\begin{figure}
\centering

\subfloat[]
{
\includegraphics[height=3.5 cm, width=4.0 cm, clip=true]{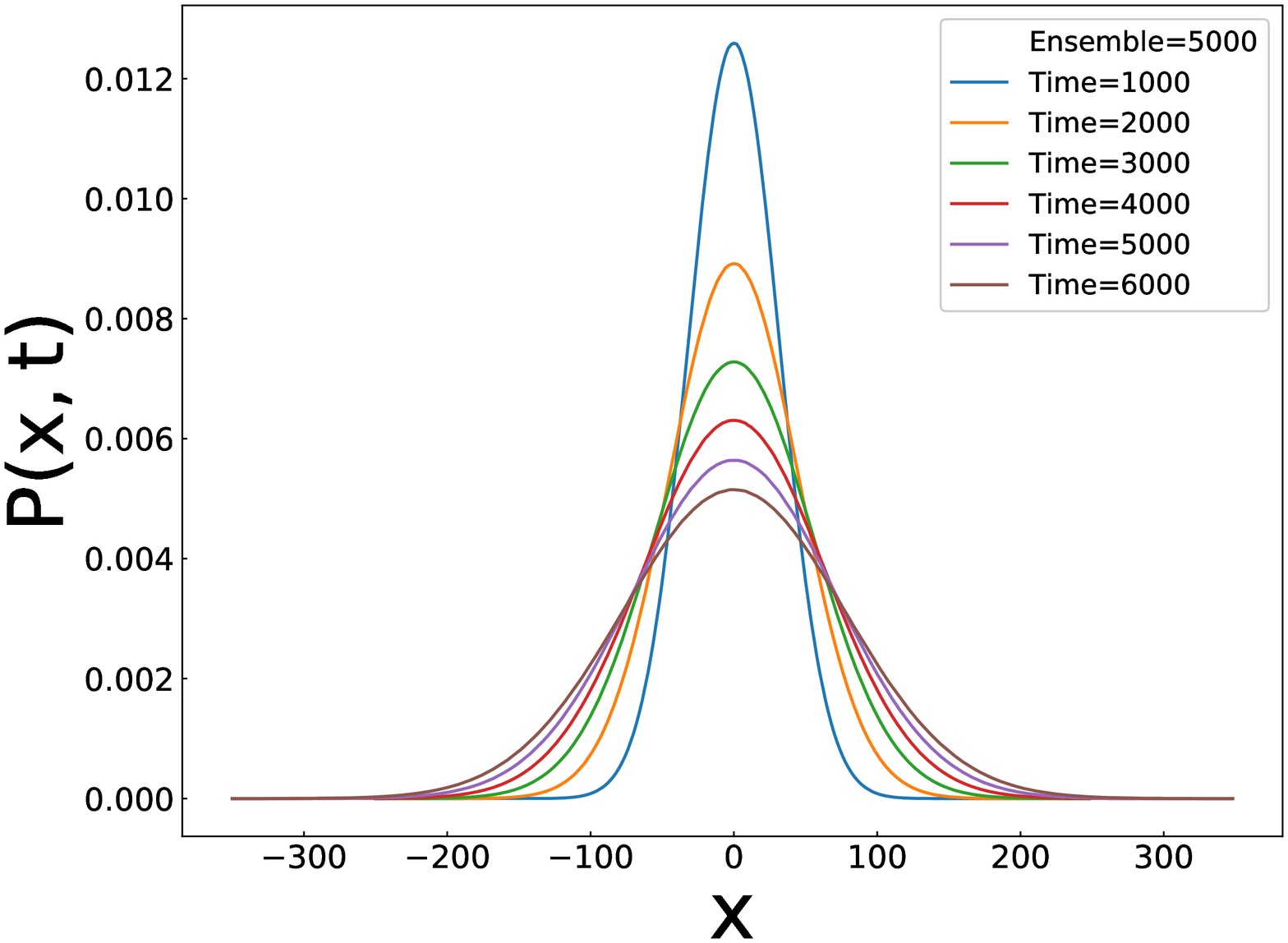}
\label{fig:3a}
}
\subfloat[]
{
\includegraphics[height=3.5 cm, width=4.0 cm, clip=true]{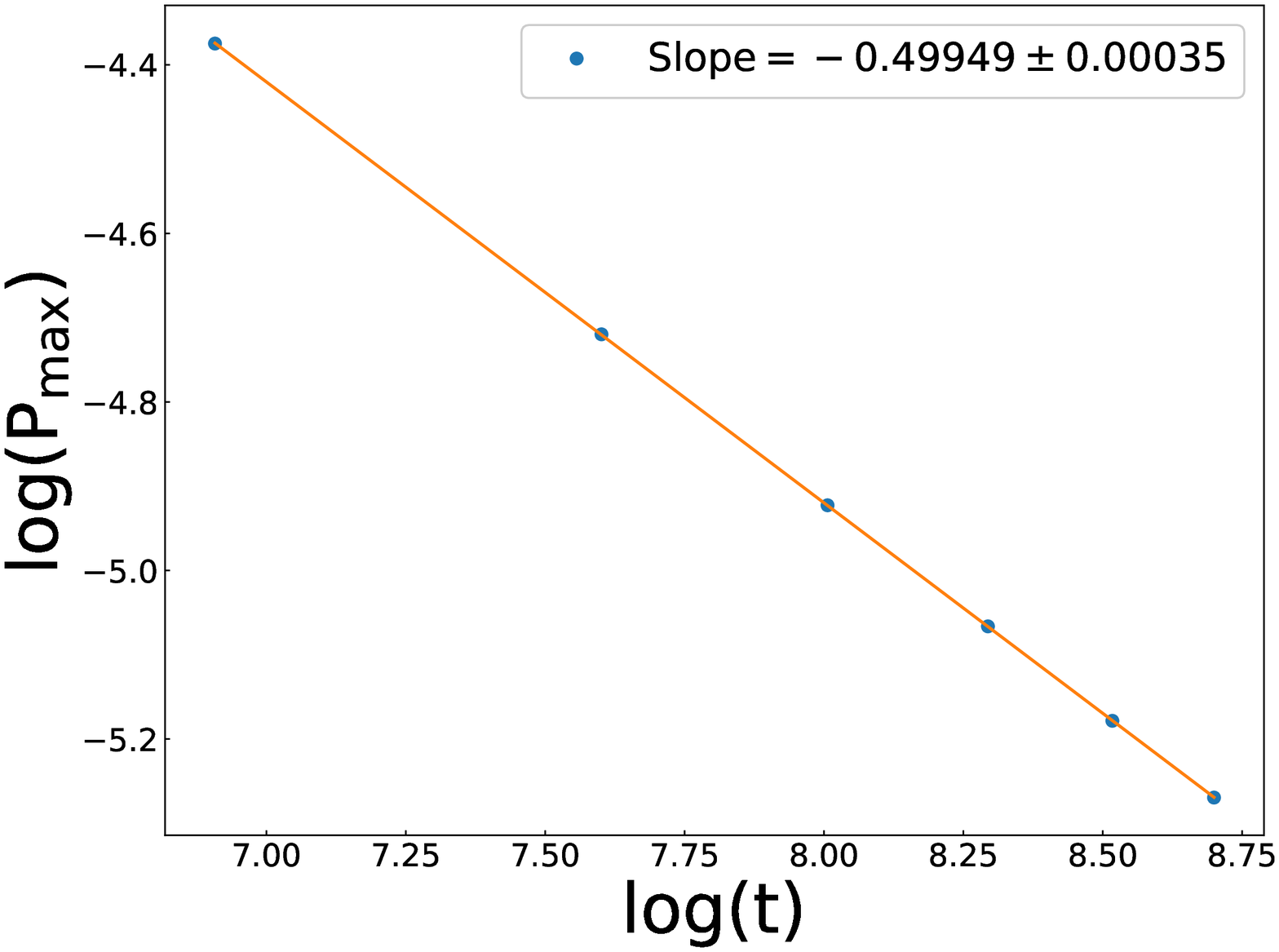}
\label{fig:3b}
}

\subfloat[]
{
\includegraphics[height=3.5 cm, width=4.0 cm, clip=true]
{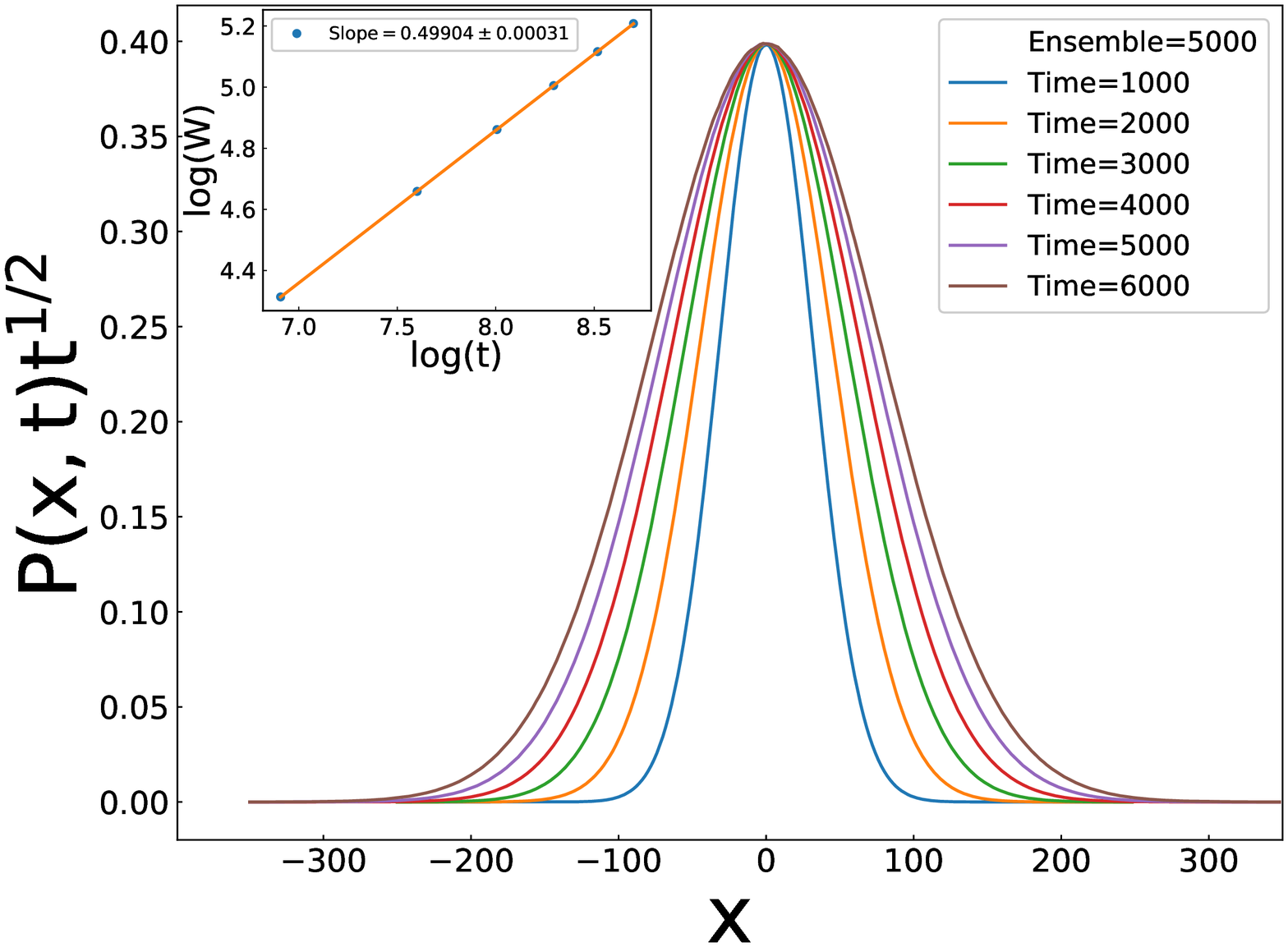}
\label{fig:3c}
}
\subfloat[]
{
\includegraphics[height=3.5 cm, width=4.0 cm, clip=true]
{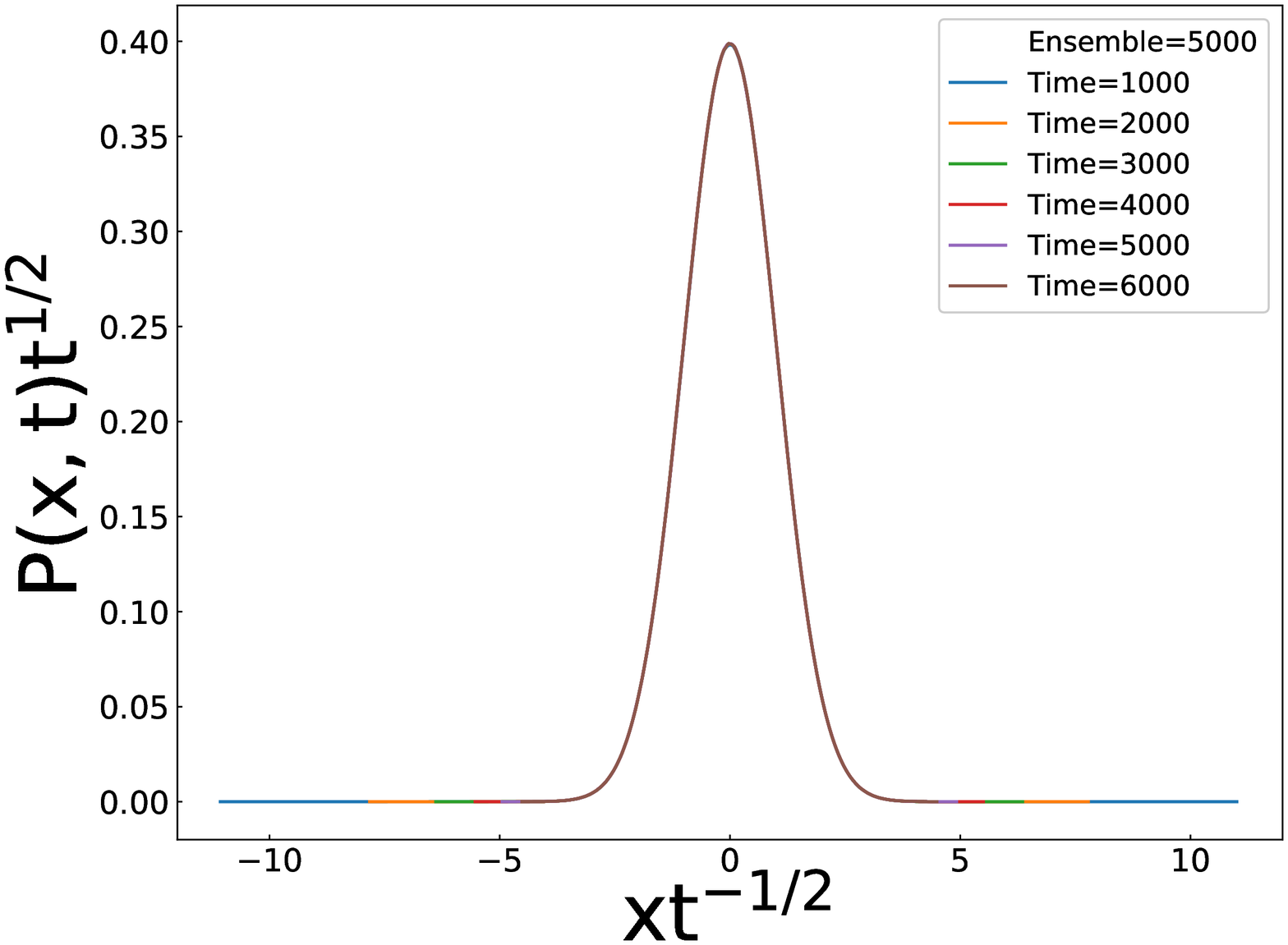}
\label{fig:3d}
}
\caption{(a) Plots of probability distribution function $P(x,t)$ as a function of $x$ for different time. (b) Peak size $P_{{\rm  max}}$ of the plots in (a) are measured for different times and the plots
of $\log(P_{{\rm max}})$ versus $\log(t)$ is shown in (b). Clearly the resulting plot is a straight line
with slope equal to $-0.5$ revealing that $P_{{\rm max}}\sim t^{-1/2}$. In (c) we plot $P(x,t)t^{1/2}$ versus $x$ and find that all the peaks
collapse at the peak. We then measure the full width at half maxima $W$ which actually represents
the root mean square displacement. In the inset we show the plot
of $\log(W)$ versus $\log(t)$ and find a straight line with slope $0.5$  which means $W\sim t^{1/2}$. Finally in (d) we plot $P(x,t)t^{1/2}$ versus $x/t^{1/2}$ and find that all the plots in 
(a) for different time collapse into one universal scaling curve.} 

\label{fig:3abcd}
\end{figure}

In Fig. (\ref{fig:3a}) we first show the plots of $P(x,t)$ as a function of position $x$ for a set of 
different time $t_i$. Note that as the walkers walk for a longer time $t=n$,
the probability of finding the walkers at a larger distance increases but this happens at the 
expense of lowering the peak value $P_{{\rm max}}$ at $x=0$ since the total area under each curve 
must always be equal to one i.e. 
$\int_{-\infty}^\infty P(x,t)dx=1$. To find out how the peak height $P_{{\rm max}}$ at $x=0$ decreases with time 
we plot $\log(P_{{\rm max}})$ versus $\log(t)$ in Fig. (\ref{fig:3b}) and find a straight line with slope
equal to $-1/2$ revealing $P_{{\rm max}}\sim t^{-1/2}$ and hence $P_{{\rm max}}t^{1/2}$ must 
be a dimensionless quantity. To prove this, we multiply the  ordinate of the 
data of $P(x,t)$ versus $x$ of Fig. (\ref{fig:3b})
by $t^{1/2}$ and re-plot the resulting it in Fig.  (\ref{fig:3c}). This is equivalent to plotting $P(x,t)t^{1/2}$ versus 
$x$ and find that all the distinct peaks of Fig. (\ref{fig:3a}) collapse superbly at $x=0$
as shown in Fig. (\ref{fig:3c}). In this way we have brought all the plots for $P(x,t)$
on equal footings as a function of $x$. We now observe that
the probability of finding the walker at larger distances increases. To find out how it increases
with time we now
measure the full width at half maximum $W$  of $P(x,t)t^{1/2}$ versus $x$ plots for different time $t$ from 
Fig. (\ref{fig:3c}). Plots 
of $\log(W)$ versus $\log(t)$ shown in the inset of Fig. (\ref{fig:3c}) results in a straight line with slope equal to $1/2$ revealing that 
$W\sim t^{1/2}$. We can thus conclude that $W$ is proportional to standard deviation $\sigma$.
In fact, we can re-write the solution in Eq. (\ref{eq:solution_difusion}) as
\begin{equation}
    \label{eq:solution_rewritten}
    P(x,t)={{1}\over{\sqrt{2\pi \sigma^2}}}\exp[-{{x^2}\over{2\sigma^2}}],
\end{equation}
where $\sigma^2=2Dt$ and standard deviation $\sigma$ is actually the root-mean square displacement.
Thus $W\sim t^{1/2}$ is consistent with our analytical solution
 given by Eq. (\ref{eq:2}). Finally, we plot  
$P(x,t)\sqrt{t}$ versus $x/\sqrt{t}$ and find that all the distinct plots of Fig. (\ref{fig:3a}) 
collapse into one universal curve as shown in Fig. (\ref{fig:3d}). Note that $W$ bear
the same dimension as that of $x$ and hence $x/\sqrt{t}$ is a dimensionless quantity.

The distribution function $P(x,t)$ for different times is distinct. However, if $P(x,t)$ and $x$ are measured 
using $t^{1/2}$ and $t^{-1/2}$ as yard-stick and then the plotting of the resulting data is equivalent to plotting  $P(x,t)\sqrt{t}$ versus $x/\sqrt{t}$. In these self-similar scaling all the data including the
data for infinitely long time walk must collapse into one universal curve which is essentially the
solution for the scaling function  
\begin{equation}
    \phi(\xi)={{1}\over{\sqrt{4\pi}}}e^{-\xi^2/4},
\end{equation}
where $\xi^2=x^2/Dt$.  Such data collapse means that random walk {\it vis-a-vis} the Brownian 
motions for different times are similar in the same sense two triangles are similar. 
Using this idea we can extrapolate data for any latter time since the plots from 
data for all time including the infinitely long time is contained in this universal curve.
Random walk or Brownian motion therefore are self-similar in nature since
walks for longer times are similar to the walkers for shorter times.

\begin{figure}
\centering

\subfloat[]
{
\includegraphics[height=3.5 cm, width=4.0 cm, clip=true]
{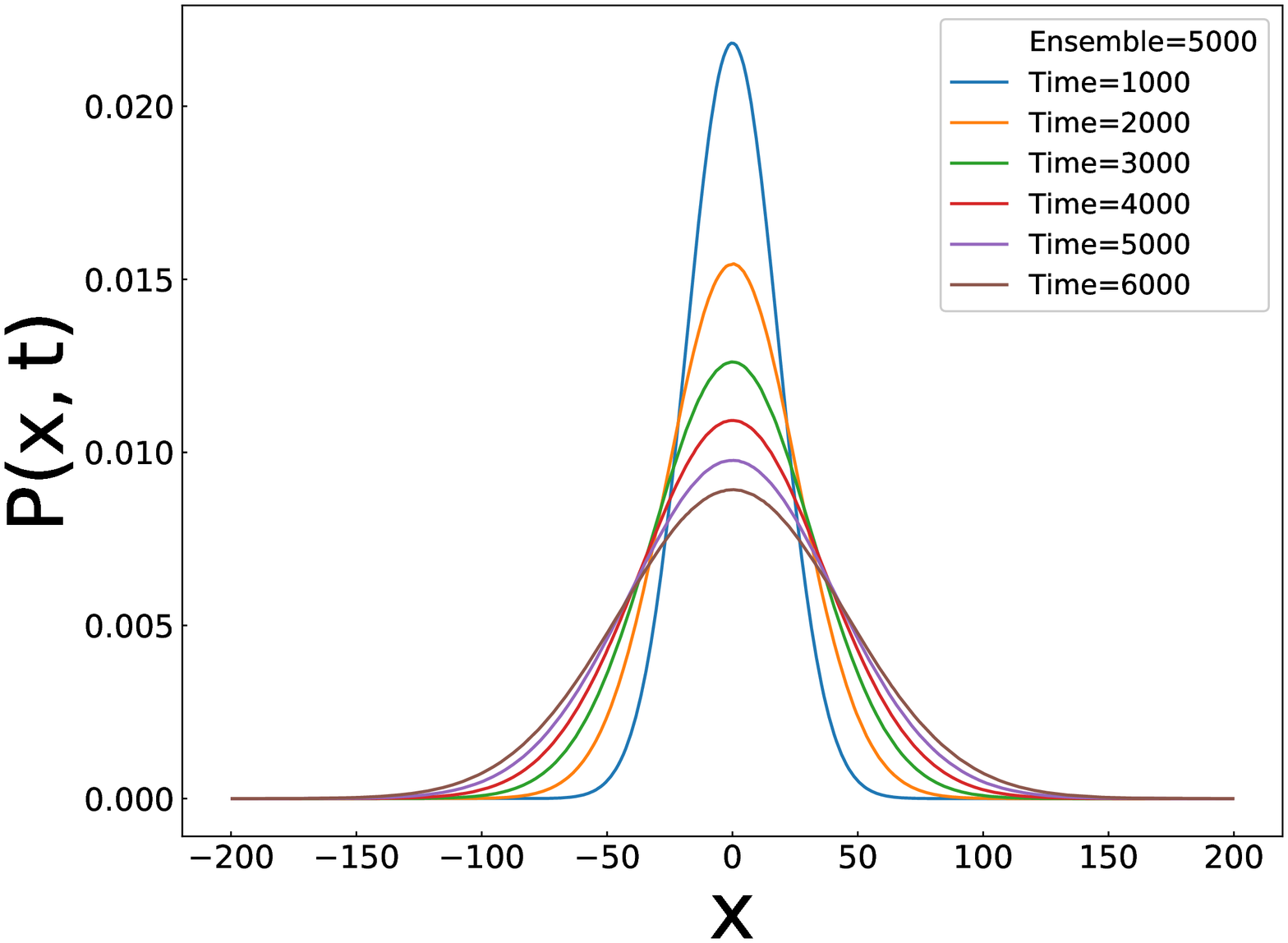}
\label{fig:4a}
}
\subfloat[]
{
\includegraphics[height=3.5 cm, width=4.0 cm, clip=true]
{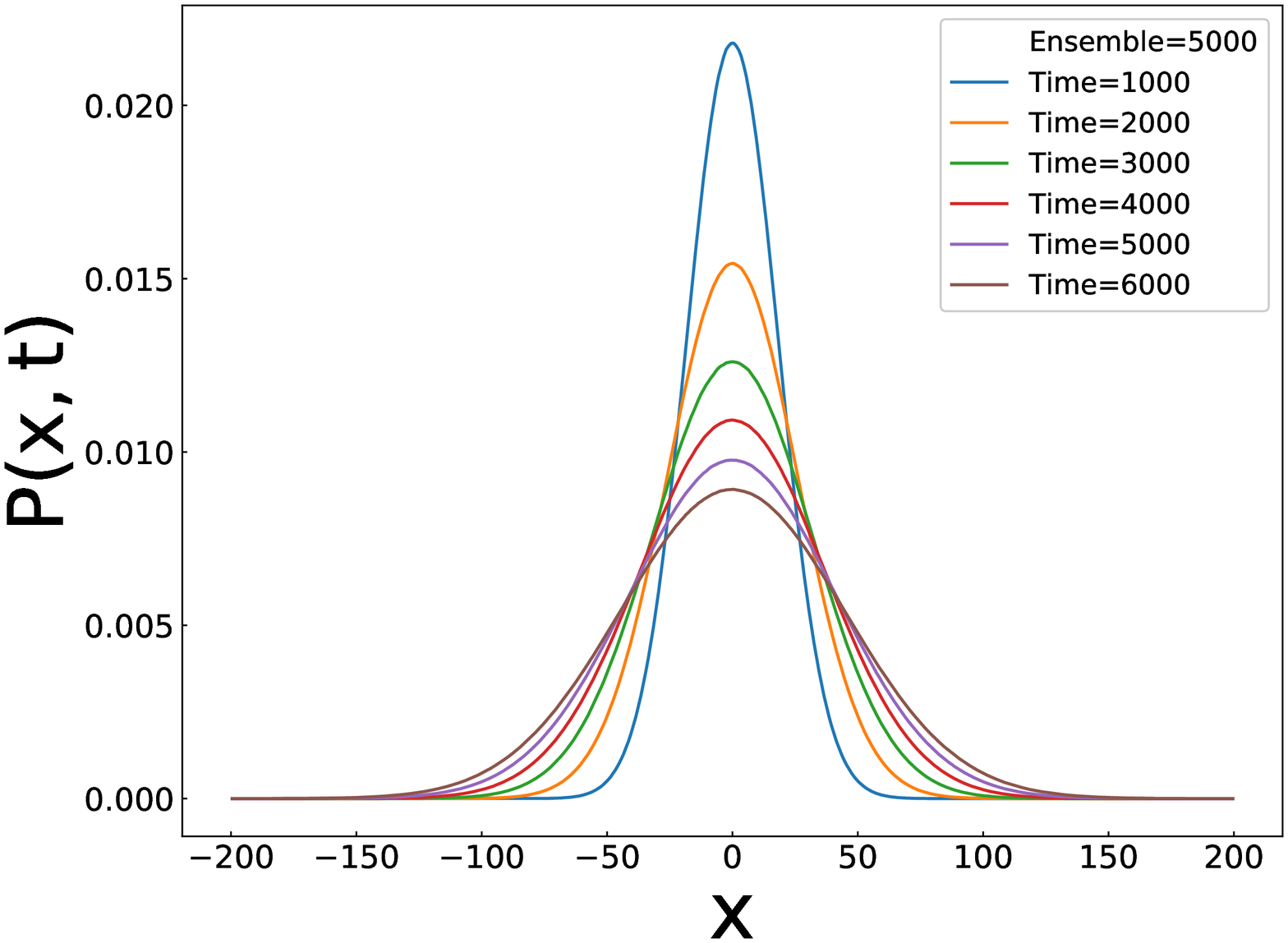}
\label{fig:4b}
}

\subfloat[]
{
\includegraphics[height=3.5 cm, width=4.0 cm, clip=true]
{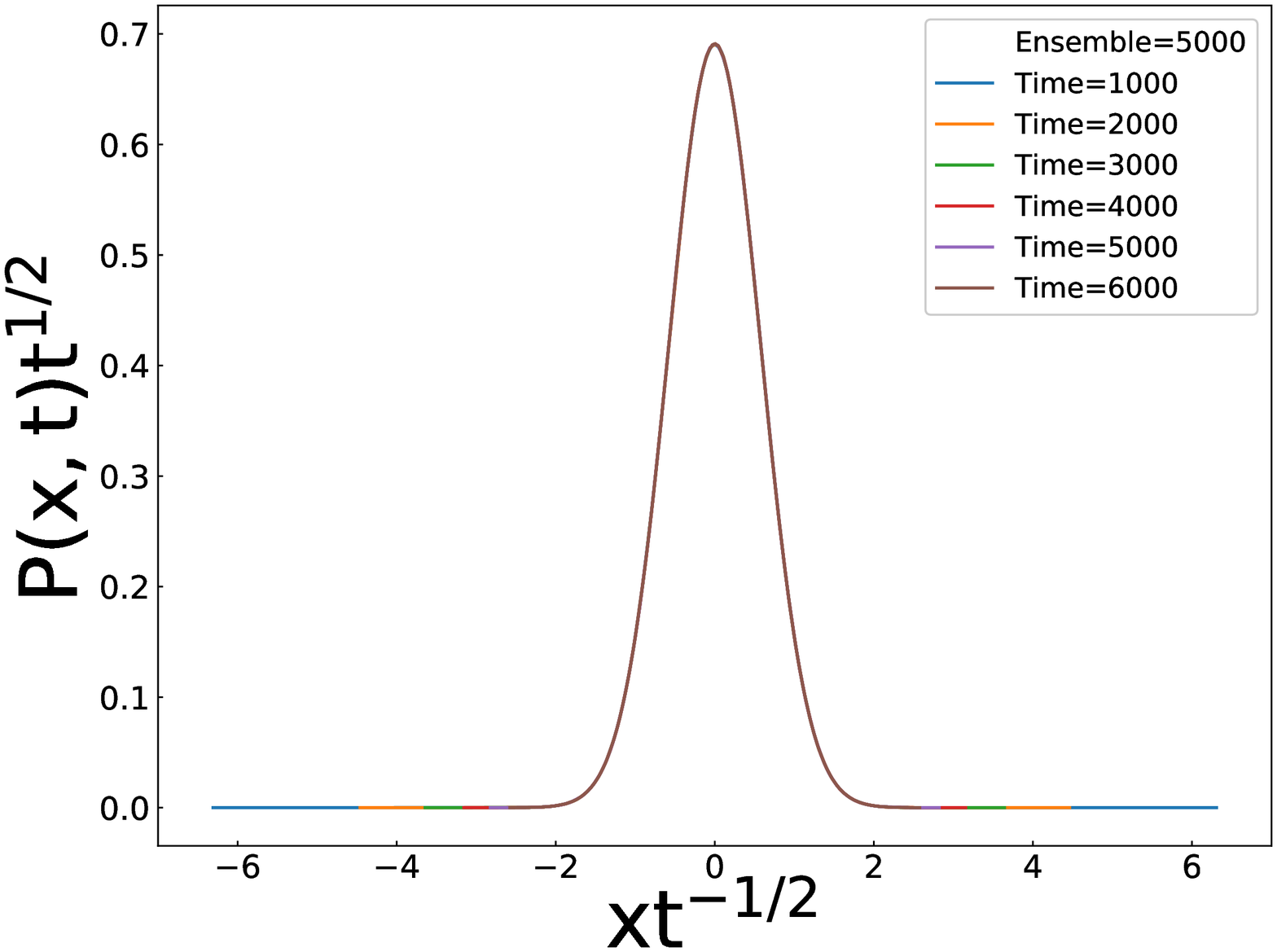}
\label{fig:4c}
}
\subfloat[]
{
\includegraphics[height=3.5 cm, width=4.0 cm, clip=true]
{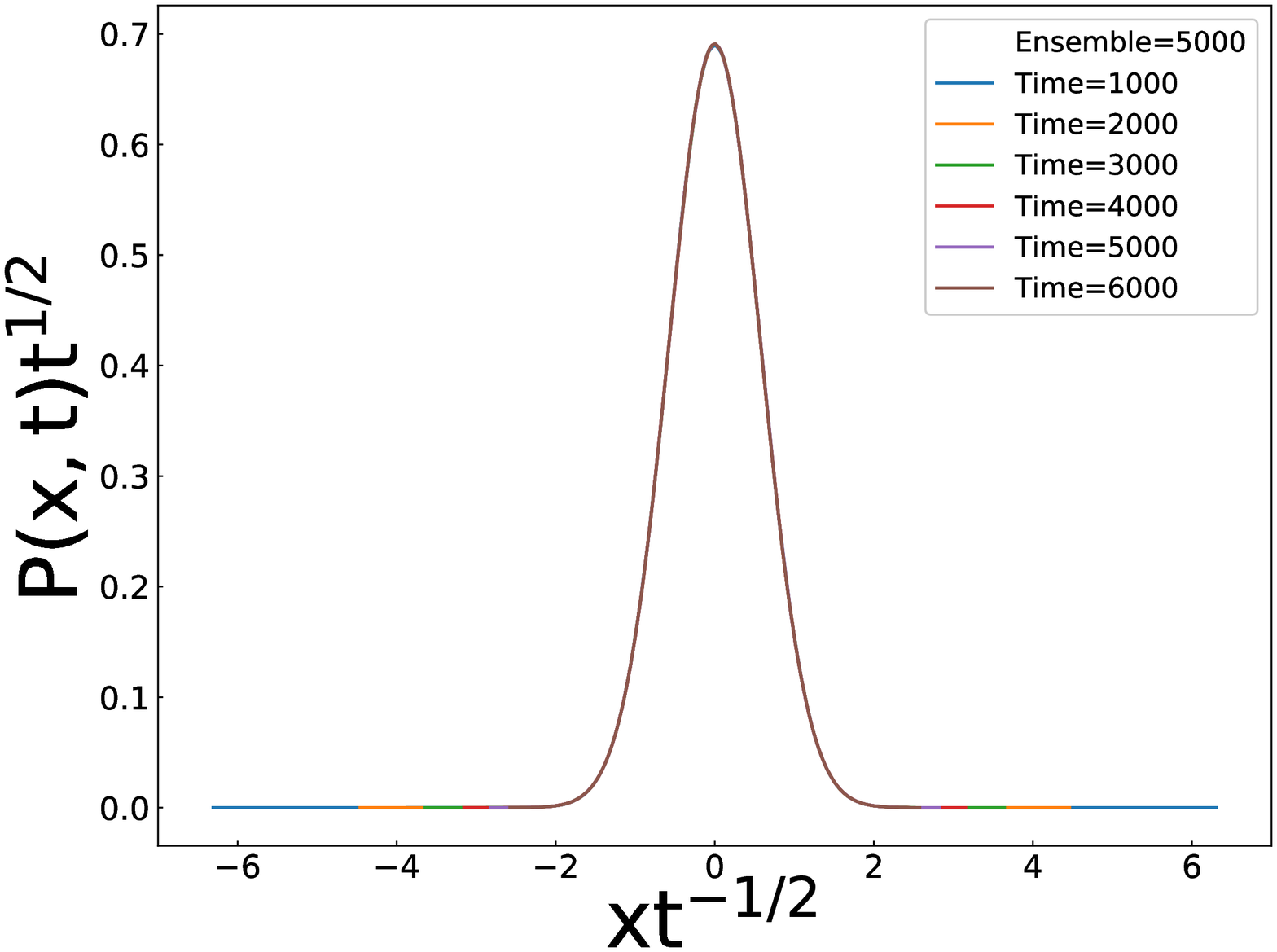}
\label{fig:4d}
}
\caption{Plots of the distribution function $P(x,t)$ versus $x$ are shown in (a) when step sizes
are random and in (b) when random step sizes are chosen in a descending order. Plots of
$P(x,t)t^{1/2}$ versus $x/t^{1/2}$ in (c) for random step size and in (d) random step sizes in a descending order. These plots clearly shows that random walk solution in either case is exactly identical to classical random walk with fixed step size apart from constant factor.} 

\label{fig:4abcd}
\end{figure}

Now the question is: What if the steps of the walkers were not of the same size but random? 
To find an answer to such
question we have performed extensive numerical simulation for a few different
situations like we have done for fixed step size. 
First, we consider the case where at each step each walker picks a random number $R$ from the
interval $[0,1]$ and the numerical value of $R$ is then chosen as the step size at that instant.
Each walker then picks another random number to decide the direction of the step in the same
way as for its classical counterpart. 
We have found that the solution for the probability distribution $P(x,t)$ is exactly
the same as it is for fixed step size (see Fig. (\ref{fig:4a})). To prove this we plot $P(x,t)\sqrt{t}$ versus $x/\sqrt{t}$ in Fig.  (\ref{fig:4c}))
and find that all the distinct plots collapse into one universal curve. It implies
that the root mean square displacement grows following Eq. (\ref{eq:rms}). 
It has far reaching
consequences as it suggests that the collision time {\it vis-a-vis} the distance travelled by 
the Brownian particles or random walkers may not be fixed, yet the solution for the distribution
function $P(x,t)$ remain the same. In fact, random walk of fixed step size
is almost impossible to find in nature. Finding random walk with random step size 
the same as that of the random walk with fixed step size 
shows how robust is the random walk problem.

\section{Random walk with shrinking step size}

\begin{figure}
\centering

\subfloat[]
{
\includegraphics[height=3.5 cm, width=4.0 cm, clip=true]
{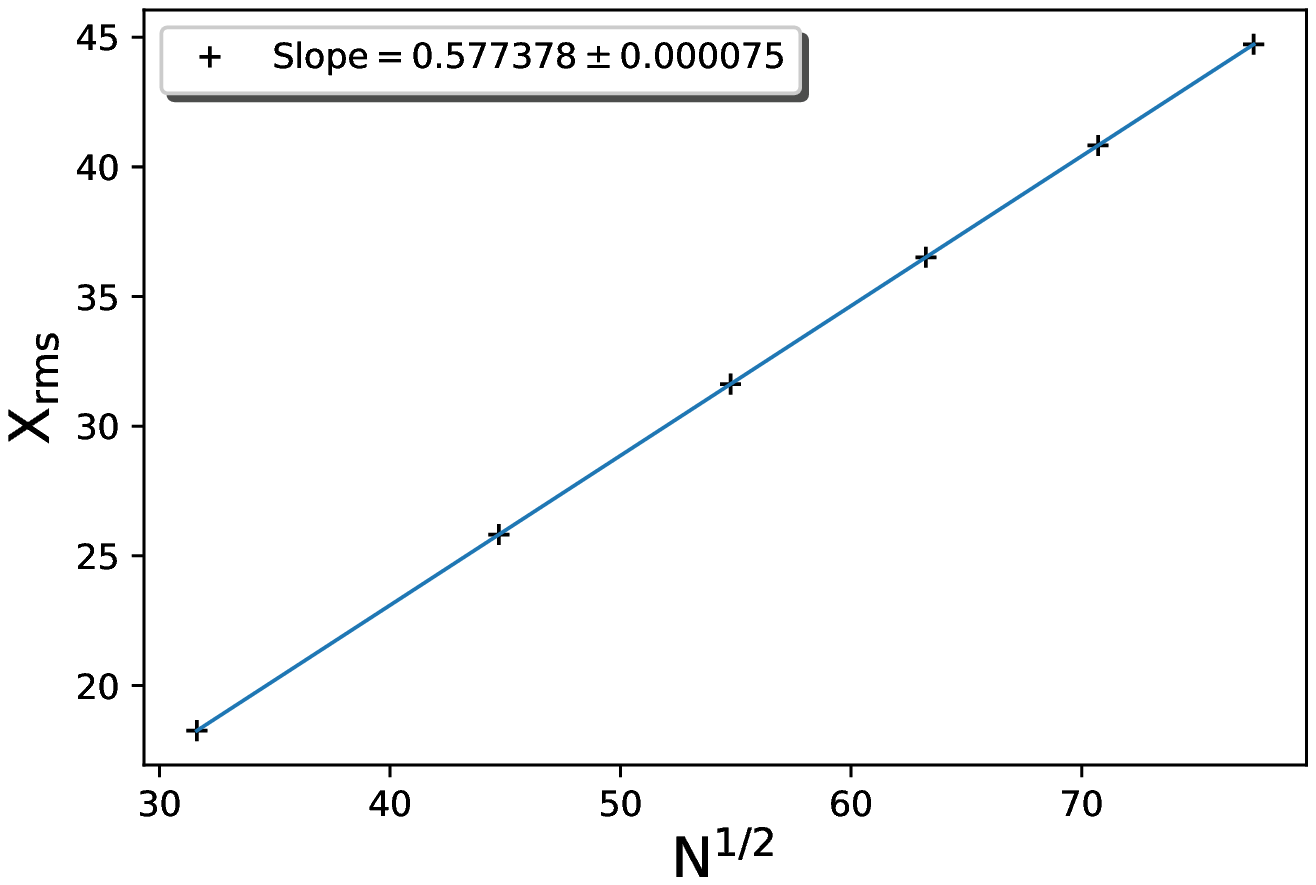}
\label{fig:5a}
}
\subfloat[]
{
\includegraphics[height=3.5 cm, width=4.0 cm, clip=true]
{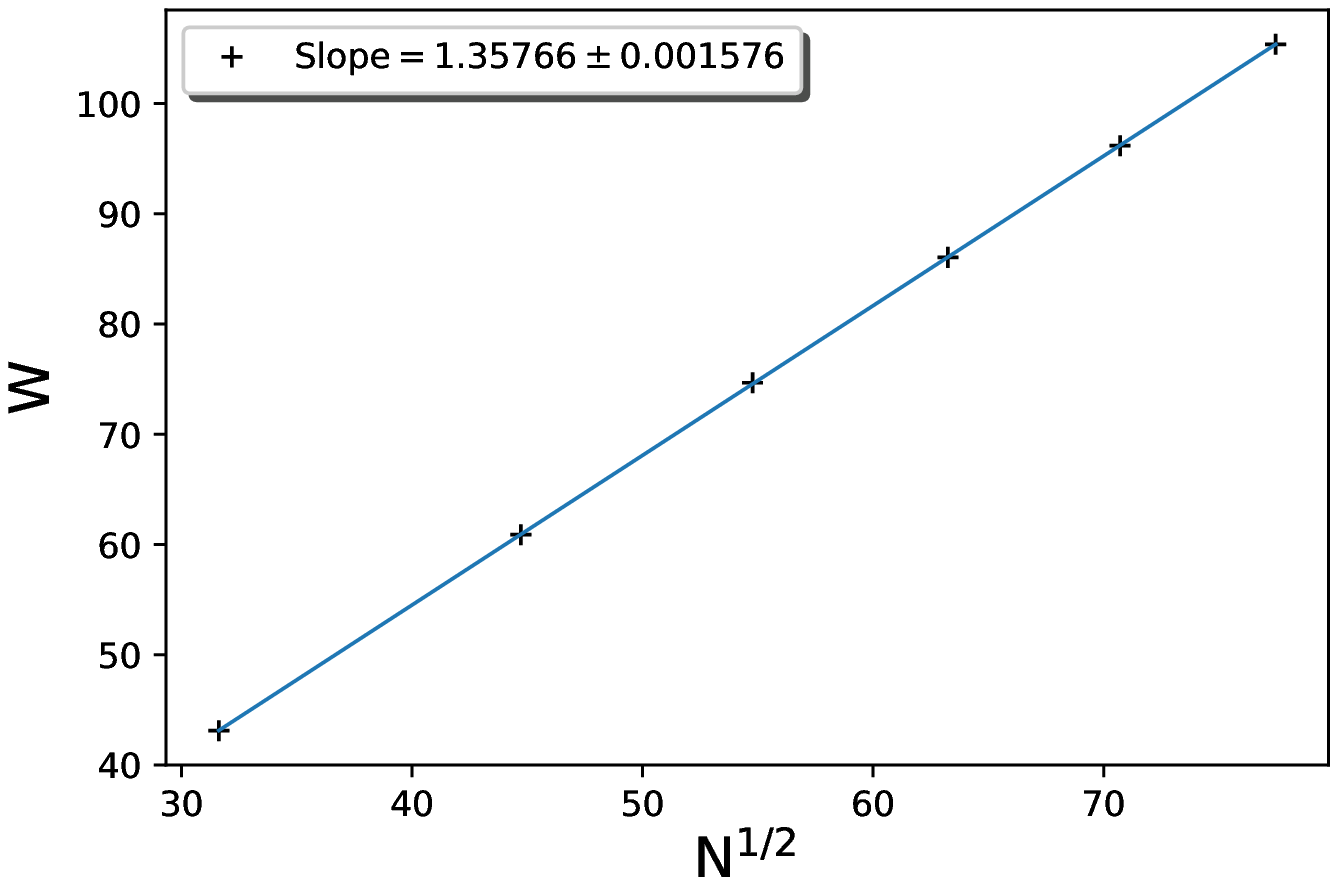}
\label{fig:5b}
}

\caption{Plots of the (a) $x_{{\rm rms}}$ versus $N^{1/2}$ and (b) 
Full width at half maximum $W$ versus $N^{1/2}$ for RW with random steps $R_n$ in 
descending order are shown and find straight lines with slopes equal to $ 0.57737$ 
and $1.3576$ respectively.} 

\label{fig:5ab}
\end{figure}

We first consider the simplest case for random walk of $N$ shrinking steps. Such problems are interesting
because of their connections to the dynamical systems \cite{ref.alexander_1, ref.alexander_2,ref.torre_shrinking}. Here, we first draw
$N$ number of random number $R_n$ uniformly from $[0,1]$ and arrange them in descending order so
that the first number is $R_1$, which is greater than the second number $R_2$ and in general
$R_i>R_{i+1}$ etc.  To create the random walk we choose the step size for the $n$th step is $R_n$
and then we choose the direction randomly. In Fig. (\ref{fig:4b}) we plot distribution function $P(x,t)$
as a function of $x$ for different walking time $t=N$. Like walk with fixed step size we find that
$P_{{\rm max}}\sim t^{-1/2}$ and the full width at half maxima $W\sim t^{1/2}$. To prove this we plot 
$P(x,t)t^{1/2}$ versus $x/t^{1/2}$ in Fig. (\ref{fig:4d}) and indeed we find all the distinct curves of Fig. (\ref{fig:4b})
collapse superbly.  Note that we have drawn $N$ random number $R_n$ from the interval $[0,1]$ with
uniform probability. Thus in the large $N$ limit, the length of the first step is $R_1=1-1/N$, 
the length of the second step is $R_2=1-2/N$ and in general the length of the $n$th step 
is $R_n = (1-n/N)$. Since the direction of steps are taken independently i.e.  
uncorrelated, the mean-square displacement after the $N$th step,
is given by:
\begin{equation}
\label{eq:msd_srw}
    \langle x^2\rangle = R_1^2+R_1^2+....+R_N^2=\sum_{n=1}^N(1-n/N)^2.
\end{equation}
In the limit $N\rightarrow \infty$ we can treat the above sum as an integral which we can 
easily integrate and find 
\begin{equation}
\label{eq:rms_srw}
       x_{{\rm rms}} =\sqrt{\langle x^2\rangle}= 0.57735\times N^{1/2}\sim t^{1/2}.
\end{equation}
We thus see that the dynamics of the random walk with shrinking steps such that the size of the $n$th 
step is $R_n$ behaves exactly the same way as for fixed and random step size. We have numerically
measured the mean square displacement $\langle x^2\rangle$ using Eq. (\ref{eq:msd_srw}) for different $N$ and plotted $\sqrt{\langle x^2\rangle}= x_{{\rm rms}}$ versus $N^{1/2}$. The resulting graph, 
as shown in Fig. (\ref{fig:5a}), is a straight line
with slope exactly at $0.57737$ supporting our result given by Eq. (\ref{eq:rms_srw}). On
the other, we find that the slope of the plots of full width at half maxima $W$ 
versus $N^{1/2}=t^{1/2}$ as shown in Fig. (\ref{fig:5b}) equals to $1.3576$. It satisfies the known relation $W=(2\sqrt{2\log2})\sigma$.

\begin{figure}
\centering

\subfloat[]
{
\includegraphics[height=3.5 cm, width=4.0 cm, clip=true]
{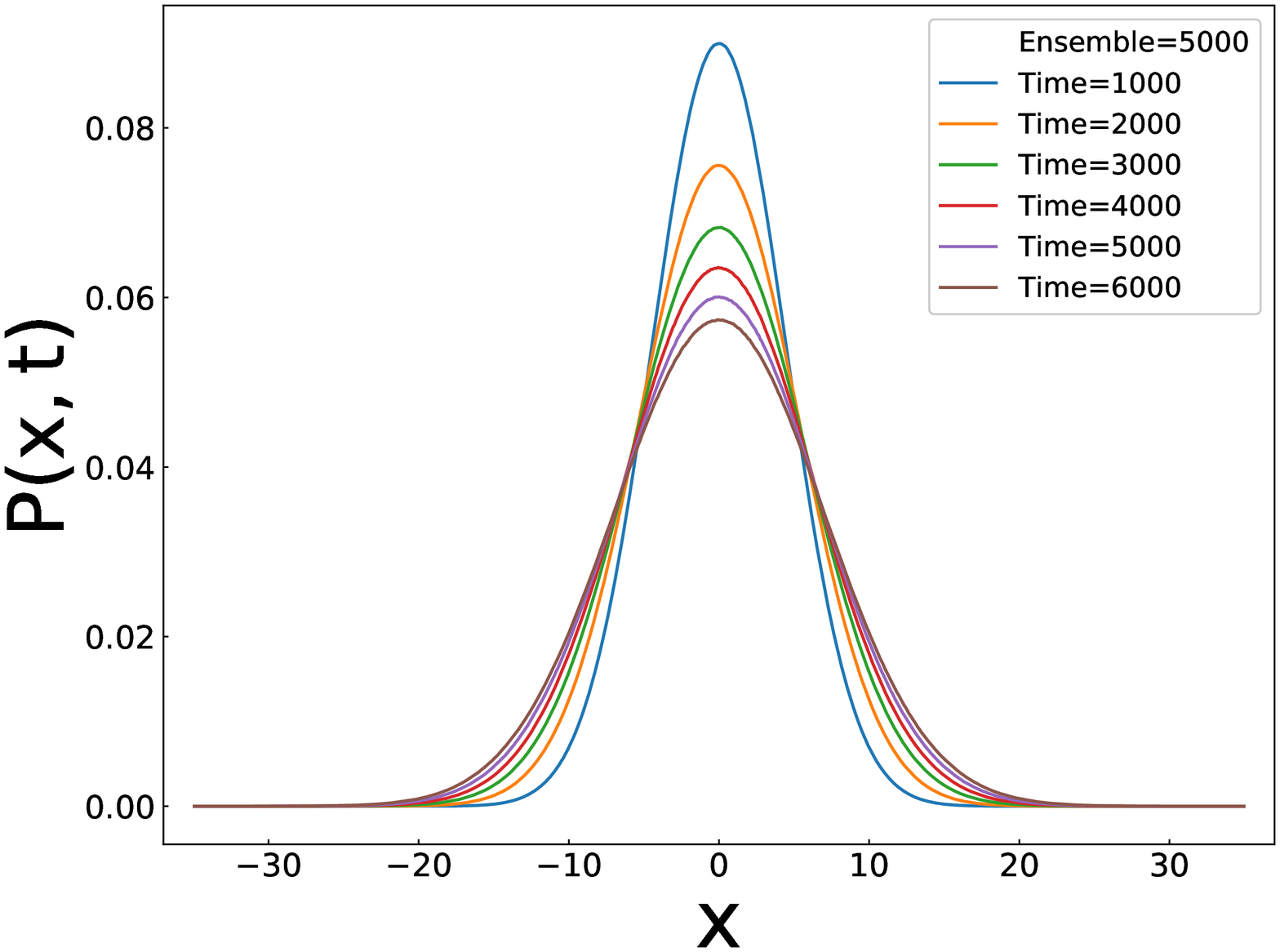}
\label{fig:6a}
}
\subfloat[]
{
\includegraphics[height=3.5 cm, width=4.0 cm, clip=true]
{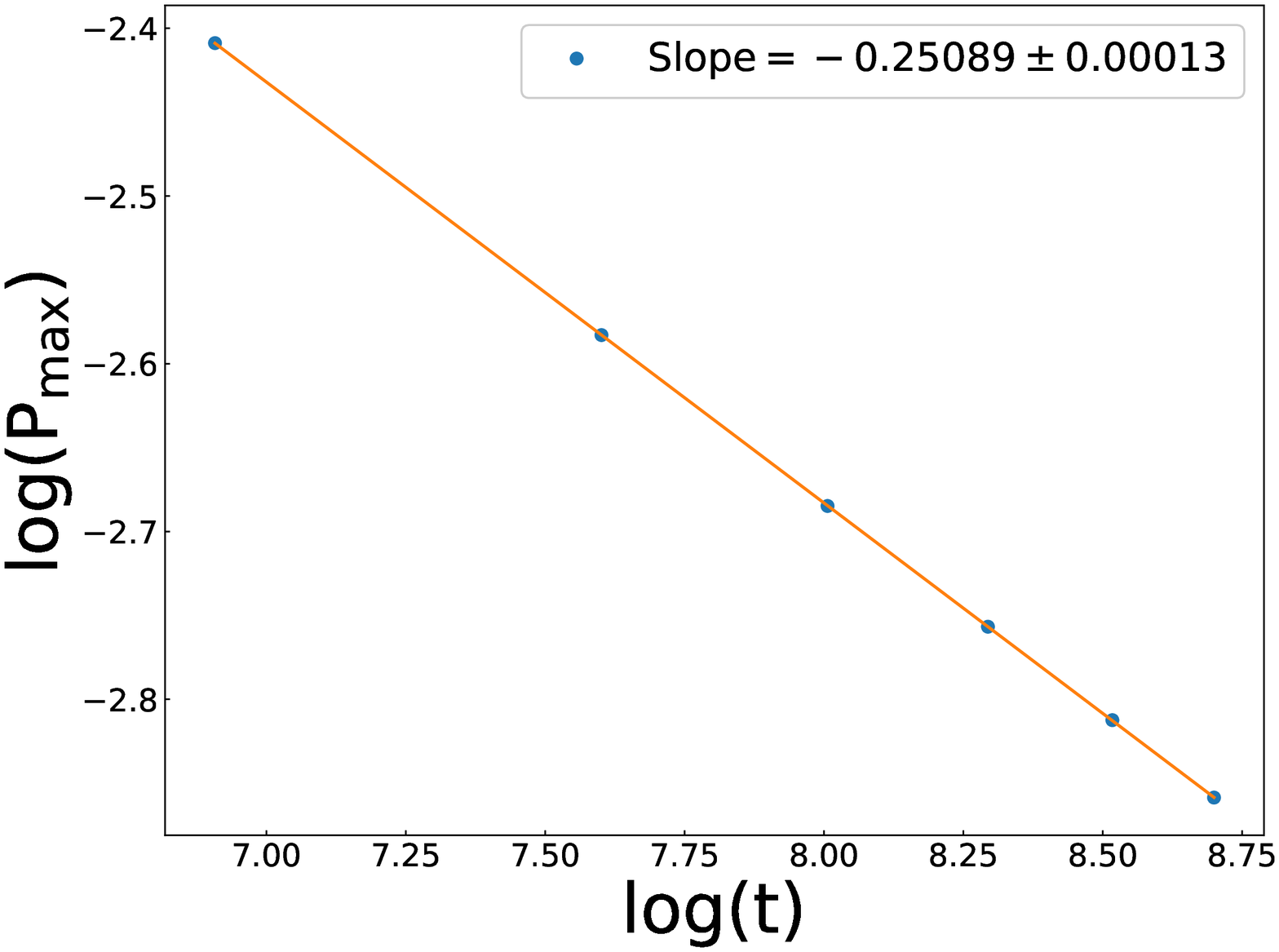}
\label{fig:6b}
}

\subfloat[]
{
\includegraphics[height=3.5 cm, width=4.0 cm, clip=true]
{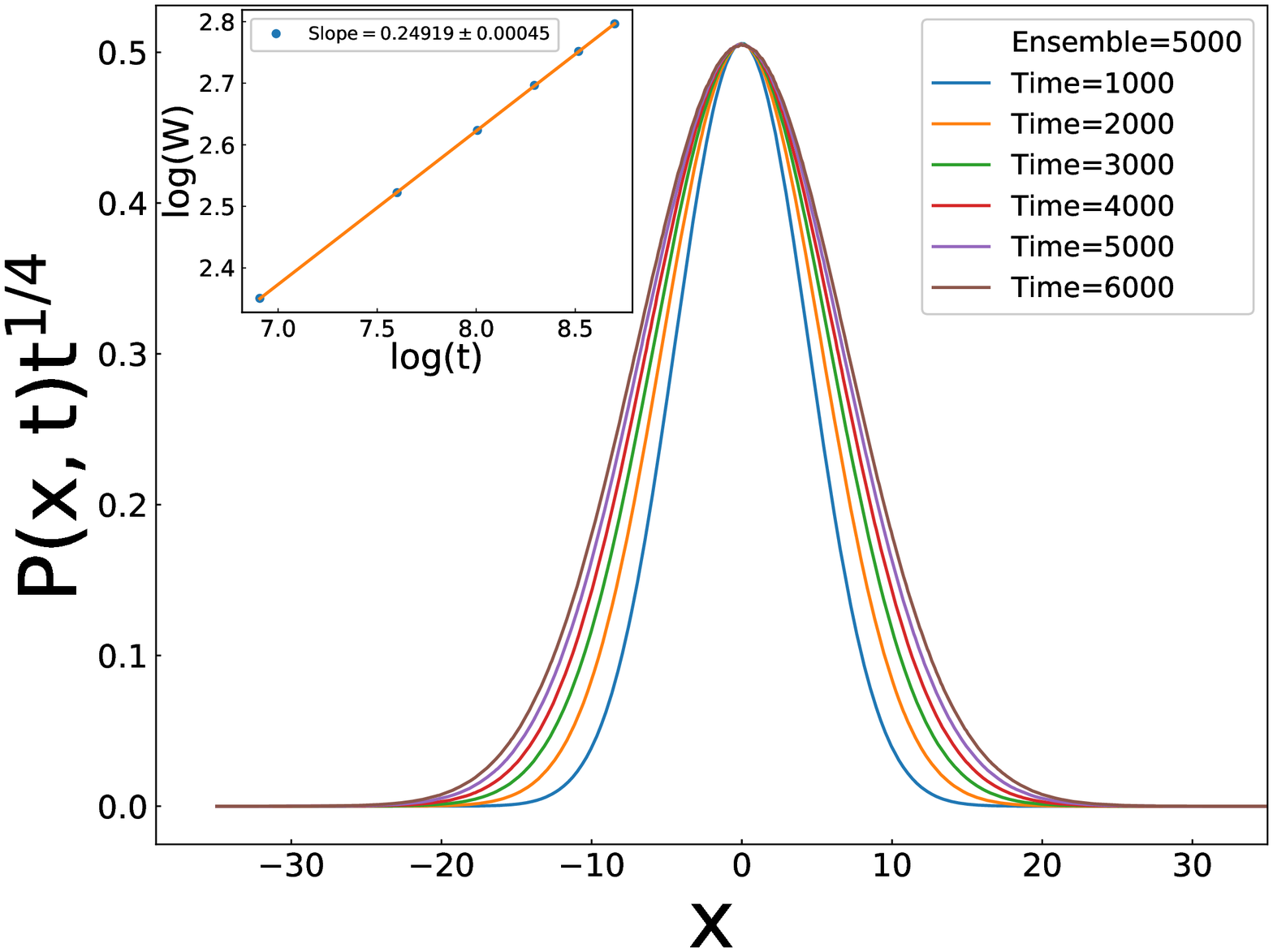}
\label{fig:6c}
}
\subfloat[]
{
\includegraphics[height=3.5 cm, width=4.0 cm, clip=true]
{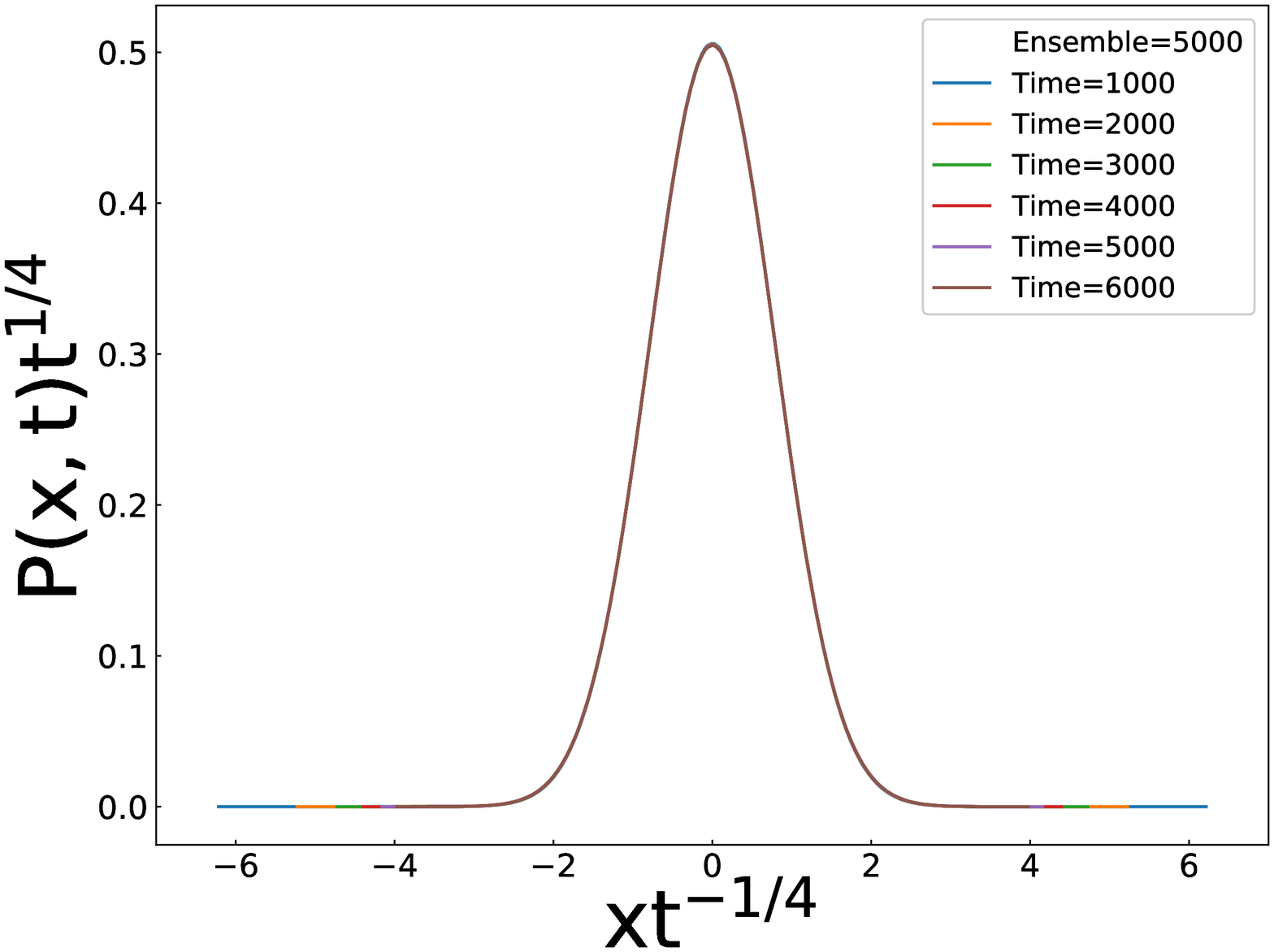}
\label{fig:6d}
}

\caption{(a) $P(x,t)$ versus $x$ plots for random walk such that the $n$th step size is equal to $S_n=R_n$
where $R_n$ is the $n$th largest random number of the $N$ random number generated within the interval $[0,1]$. (b) Peak height $P_{{\rm  max}}$ of the plots in (a) are measured for different times and in (b) we  plot of $\log(P_{{\rm max}})$ versus $\log(t)$. 
Clearly the resulting plot is a straight line
with nontrivial slope equal to $-0.25$ revealing that $P_{{\rm max}}\sim t^{-1/4}$. In (c) we plot $P(x,t)t^{1/4}$ versus $x$ and find that all the peaks
collapse at the peak. The full width at half maxima $W$ are then measured and in the inset  the plot
of $\log(W)$ versus $\log(t)$ are drawn. We find a straight line with slope $0.25$ which means $W\sim t^{1/4}$. Finally in (d) we plot $P(x,t)t^{1/4}$ versus $x/t^{1/4}$ and find that all the plots in (a) for different time collapse into one universal scaling curve.} 

\label{fig:6abcd}
\end{figure}

In 2003 Krapivsky and Redner studied a few interesting variants of the random walk problem \cite{ref.redner_shrinking}.
They considered the case in which the length of the $n$th step
\begin{equation}
\label{eq:shrinking_size}
    S_n=\lambda^n,
\end{equation} 
where $\lambda<1$ and it is assumed to be a fixed value across the whole journey. 
They found that the support of the distribution function $P(x,t)$ is a Cantor set for $\lambda<1/2$.
However, for $1/2\leq \lambda <1$ there is countable infinite set of $\lambda$ values for which
$P(x,t)$ is singular. Of all the $\lambda$ values, one of the strikingly interesting 
results have been found if one chooses for $\lambda$ equal to inverse golden number $(\sqrt{5}-1)/2$. 
We shall now study the case where the $n$th step size $S_n$ is given by Eq. (\ref{eq:shrinking_size})
except now we choose $\lambda$ not a fixed number rather $\lambda=R_n$
where $R_n$s are random numbers picked from the interval $[0,1]$ and arranged in a descending order such that $R_1>R_2>R_3>.....>R_N$. This time we find interesting
results which are significantly different from all known cases including
the work of Krapivsky and Redner \cite{ref.redner_shrinking}.

\begin{figure}
\centering

\subfloat[]
{
\includegraphics[height=3.5 cm, width=4.0 cm, clip=true]
{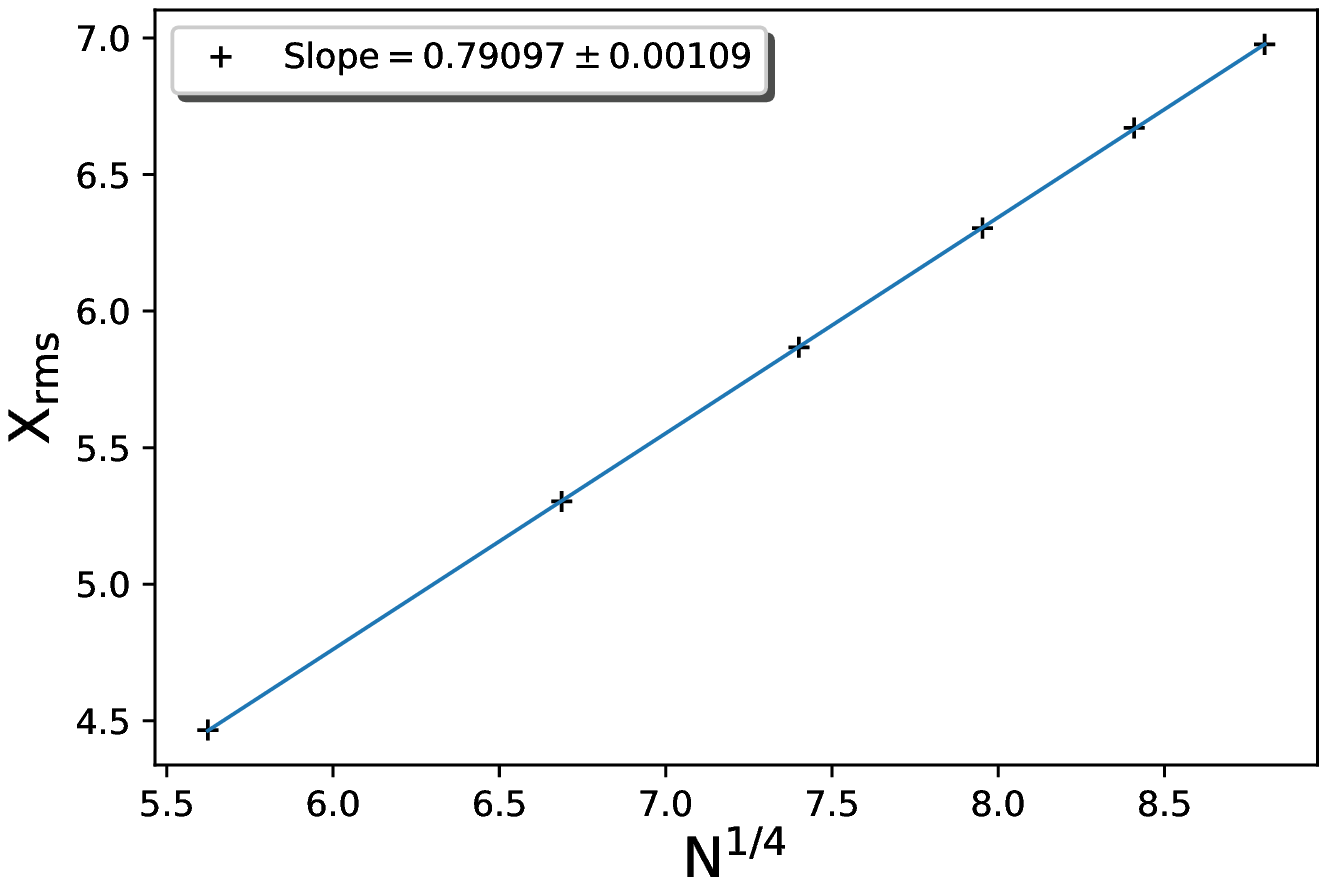}
\label{fig:7a}
}
\subfloat[]
{
\includegraphics[height=3.5 cm, width=4.0 cm, clip=true]
{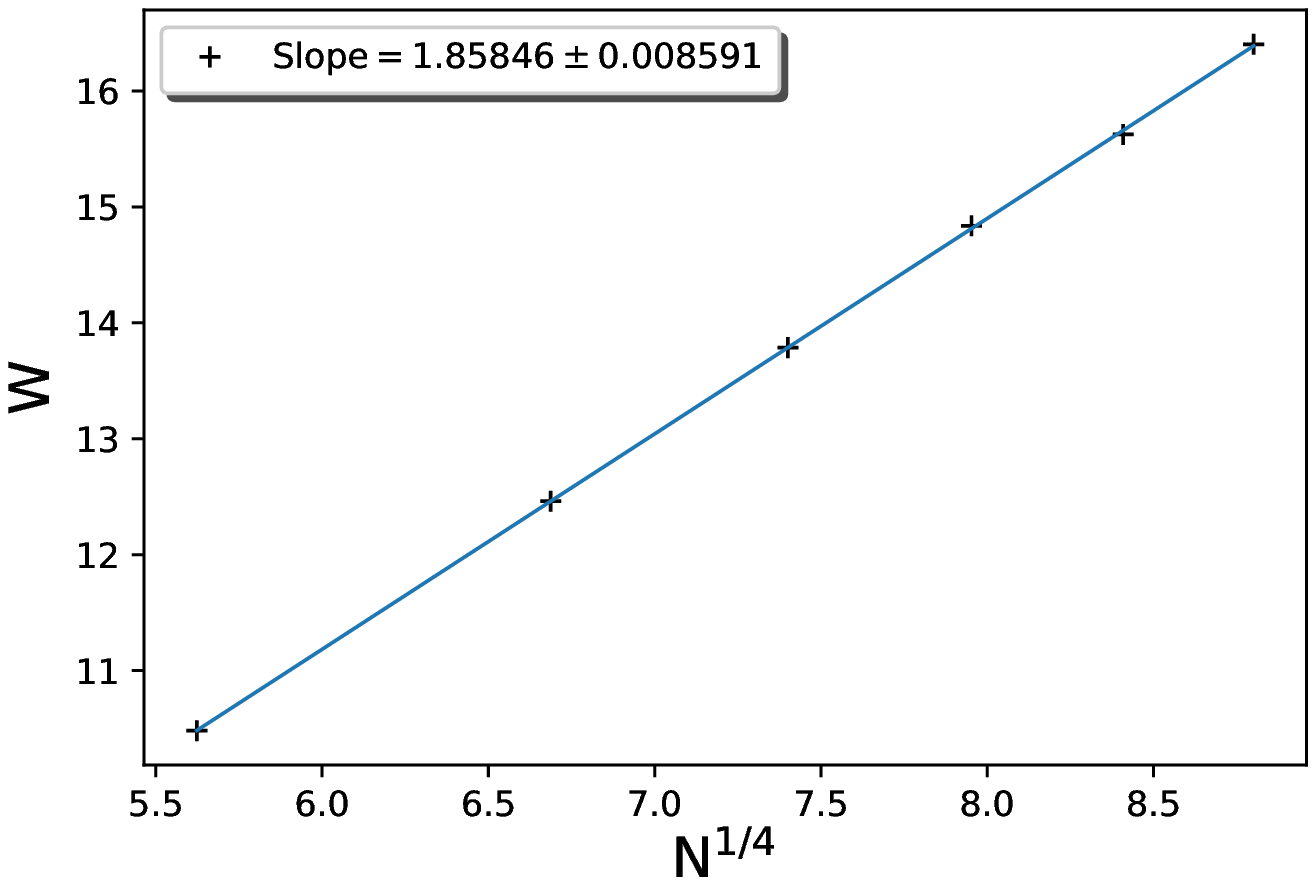}
\label{fig:7b}
}

\caption{Plots of the (a) $x_{{\rm rms}}$ versus $N^{1/4}$ and (b) $W$ versus $N^{1/4}$ for RW with 
shrinking steps $R_n^n$ and find straight lines with slopes equal to $ 0.791$ and $1.86$ respectively.} 

\label{fig:7ab}
\end{figure}

In Fig. (\ref{fig:6a}) we show plots of $P(x,t)$ versus
$x$ for different times. It is clear from the figures that the shape of the distribution function 
curve is different from all the cases where $\lambda$ assumes a fixed value \cite{ref.redner_shrinking}.
 It is actually more similar to the classic random walk problem, albeit the
 exponents are different, than the works of Krapivsky and Redner. 
Note that in the case of constant $\lambda$, the value
of $\lambda^n$ decreases with increasing $n$ if $\lambda<1$. In our case, we have first
generated $t$ number of random number from the interval $[0,1]$ and then
arranged them in a descending order so that the $n$th smallest number is $R_n$. 
We then choose the size of the $n$th step of the walker as $R_n^n$ and hence 
the larger the $n$ value the smaller the step size. Interestingly, we find that
the probability distribution function $P(x,t)$ looks similar to that of the classical random
walk problem. Nevertheless, we then measure the peak height $P_{{\rm max}}$ as a function
time $t$. We then plot $\log(P_{{\rm max}})$ versus $\log (t)$ in (\ref{fig:6b}) and find that it
yields a straight line with slope $-0.25$ instead of $0.5$ for its classical counterpart. 
In Fig. (\ref{fig:6c}) we now plot $P(x,t)t^{0.25}$ versus $x$ and find that all the peaks
of Fig. (\ref{fig:6a}) collapse at $x=0$. We now measure the full width
at the half maxima $W$ of Fig. (\ref{fig:6c}). Plotting
$\log(W)$ versus $\log(t)$ in the inset of Fig. (\ref{fig:6c}) once again gives straight line but with slope $0.25$. It implies that 
root-mean square displacement $x_{{\rm rms}}$ increases like $t^{1/4}$. We  now plot
$t^{0.25}P(x,t)$ versus $xt^{-0.25}$ in Fig. (\ref{fig:6d}) and find that all the distinct plots of Fig. (\ref{fig:6a})
collapse into one universal curve. Thus, the random walk with shrinking step size, so that 
the step size of the $n$th step size equals to $R_n^n$, exhibits the same self-similar solution as
it is for the fixed step size except the exponents are $0.25$ instead of $0.5$.

To understand why RW with $S_n=R_n^n$ is different from that $S_n=R_n$ we find the root mean square displacement and see how it behaves with time $t$. The length of the $n$th step is now 
\begin{equation}
   R_n=(1-n/N)^n,    
\end{equation}
which in the limit $N\rightarrow \infty$ can be re-written as
\begin{equation}
   R_n= e^{n\log (1-n/N)}\approx e^{-n^2/N}.    
\end{equation}
We thus see that the step size decreases exponentially with $n$ where in the case of $S_n=R_n$ it decreases linearly with $n$. We also observe that when $n<\sqrt{N}$ the step size is of the order of one but
when $n>\sqrt{N}$ the step size quickly becomes vanishingly small. Like before we can again calculate
the root mean square displacement
\begin{equation}
    \langle x^2\rangle = R_1^2+R_1^2+....+R_N^2,
\end{equation}
where $R_1>R_2......>R_N$. In the large $N$ limit we can write it as
\begin{eqnarray}
\langle x^2\rangle &=& \sum_{n=1}^N(1-n/N)^{2n}, \\ \nonumber
    &=& \sum_{n=1}^N\Big ((1-n/N)^N\Big)^{2n/N}\approx\int_1^N e^{-2n^2/N}dn,
\end{eqnarray}
and in the limit $N\rightarrow \infty$ we can write it as
\begin{equation}
     \langle x^2\rangle=\sqrt{{{N}\over{2}}}\int_0^\infty e^{-x^2} dx={{\sqrt{\pi}}\over{2^{3/2}}}
N^{1/2}.
\end{equation}
Thus, the width of the distribution $P(x,N)$ grows as 
\begin{equation}
       x_{{\rm rms}} =\sqrt{\langle x^2\rangle}=0.79162 \times N^{1/4}=0.79162\times t^{1/4},
\end{equation}
which clearly support our numerical findings. 
Plotting $x_{{\rm rms}}$ versus $N^{1/4}$ we in fact find that slope is almost
equal to $0.79162$ (see Fig. (\ref{fig:7a}). On the other hand, plotting of $W$ versus $N^{1/4}$ 
in Fig. (\ref{fig:7b}) gives the
slope equal to $1.86$ which once again is consistent with the fact that 
$W=(2\sqrt{2\log 2})\sigma$.

\section{Discussion}

We have given a comprehensive description of the most studied random walk problem 
focusing mainly on its self-similar property and verifying the results numerically. 
We also highlighted its connection to diffusion processes. It is noteworthy  
to mention that diffusion is a ubiquitous phenomena and knowing its connection to random walk helps
us to simulate diffusion in the computer. 
At first, we have discussed the idea of similarity and self-similarity {\it vis-a-vis} the dynamic scaling
and its deep connection to Buckingham $\Pi$ theorem where the notion of dimensionless quantity 
plays a significant role. Using the Markov chain identity it has been shown that the dynamics of the random walk problem is
governed by the diffusion equation. We then used the idea of Buckingham $\Pi$ theorem to obtain solution
of the diffusion equation as it provides deep insight into the problem. One of the primary goals of this
work is to perform extensive numerical solution to verify the analytical solution. Besides,
we show that the random walk problem is robust with respect to step size albeit up to some extent.
On the other hand, Krapivsky and Redner studied random walks with geometrically shrinking steps, 
in which the size of the $n$th step is considered to be $\lambda^n$ with $\lambda<l$. In particular
they choose $\lambda=2^{-m}$ where $m=1,2,3$ etc. for which analytical solution is possible and
it is non-trivial. Furthermore, they choose $\lambda=(\sqrt{5}-1)/2$ and found highly non-trivial 
self-similar features. 

We have chosen $\lambda$ equal to a random number within an interval $[0,1]$ instead of a fixed
number.
In contrast, we have also studied two variants of the random walk with shrinking step. First, we have 
chosen the $n$th step size equal to $R_n$ such that $R_1>R_2>R_3>....>R_{N-1}>R_N$ where $R_n$s
are random numbers drawn from the interval $[0,1]$. In this case, we have found that the results
are exactly the same as for classical fixed step size random walk. Second, we have chosen 
 shrinking step size so that the $n$th step size equal to $R_n^n$ instead of $R_n$. This is similar to
 the geometric random walk of Krapivsky and Redner except the fact that they choose a constant value for
 $R_n$. Interestingly, results too are very different. We found that the overall features of random
 walk, such that the $n$th step size $R_n^n$, are the same in the sense that the distribution function
 $P(x,t)$ are still Gaussian and they still obey dynamic scaling. However, the peak height 
 $P(x,t)$ decays like $t^{-1/4}$ instead of $t^{-1/2}$ and root mean square displacement increases like
 $t^{1/4}$ instead of $t^{1/2}$. It would be interesting to see what happens if we extend the present
 work in higher dimension which we intend to do in our future endeavor. 
 
MKH would like to thank Professor Sidney Redner for critical reading of the manuscript and 
his valuable comments especially in helping to find analytical argument for random walk with 
algebraically shrinking steps.


\begin{thebibliography}{99}

\bibitem{ref.barenblatt} G. I. Barenblatt, Scaling, Self-similarity, and 
Intermediate Asymptotics (Cmpridge University Press, 1996).
\bibitem{ref.hassan_santo} S. Banerjee, M. K. Hassan, S. Mukherjee  and  A Gowrisankar, {\it 
Fractal Patterns in Nonlinear Dynamics and Applications} (CRS press,
Tayor \& Francis group, New York, 2020).
\bibitem{ref.redner_krapivsky} P. L. Krapivsky, S. Redner and E. Ben-Naim, {\it A Kinetic View of Statistical Physics} (Cambridge University Press, New York, 2010).
\bibitem{ref.Hassan_Rahman_1} M. K. Hassan and M. M. Rahman, Phys. Rev. E {\bf 92} 040101(R) (2015); 
{\it ibid}  {\bf 94} 042109 (2016).
\bibitem{ref.hassan_didar} M. K. Hassan, D. Alam, Z. I. Jitu and M. M. Rahman, Phys. Rev. E, {\bf 96} 050101(R) (2017).
\bibitem{ref.hassan_sabbir} M. M. H. Sabbir and M. K. Hassan, Phys. Rev. E {\bf 97} 050102(R) (2018).
\bibitem{ref.pearson} K. Pearson,  Nature {\bf 72} 294 (1905).
\bibitem{ref.rayleigh} L. Rayleigh, Nature {\bf 72} 318 (1905).
\bibitem{ref.einstein} A. Einstein, Ann. Phys. {\bf 19} 371  (1906). 
\bibitem{ref.smoluchowski} M. Smoluchowski, Phys. Zeit {\bf 17} 557 (1916).
\bibitem{ref.jessen} B. Jessen and A. Wintner, Trans. Am. Math. Soc. {\bf 38} 48 (1935) 
\bibitem{ref.kershner} B. Kershner and A. Wintner,  Am. J. Math. {\bf 57} 541 (1935). 
\bibitem{ref.winter} A. Wintner,  Am. J. Math. {\bf 57} 827 (1935).
\bibitem{ref.erdos}  P.  Erd\"{o}s,  Am. J. Math. {\bf 61} 974 (1939);  {\bf 62} 180 (1940).
\bibitem{ref.garsia} A. M. Garsia,  Trans. Am. Math. Soc. {\bf 102} 409 (1962);  Pac. J. Math. {\bf 13} 1159 (1963).
\bibitem{ref.barkai} E. Barkai and R. Silbey, Chem. Phys. Lett. {\bf 310} 287 (1999); Phys. Chem. B, {\bf 104} 342 (2000).
\bibitem{ref.h_weiss} G. H. Weiss and J. E Kiefer, J. Phys. A {\bf 16} 489 (1983).
\bibitem{ref.hassan_dynamic_scaling} M. K. Hassan, M. Z. Hassan and N. I. Pavel, J. Phys. A: Math. Theor. {\bf 44} 175101 (2011).
\bibitem{ref.hassan_liana} M. K. Hassan, L. Islam, S. A. Haque, Physica A {\bf 469} 23 (2017).
\bibitem{ref.hassan_liana_debashish} M. K. Hassan, L. Islam, S. A. Haque, Physica 
A {\bf 469} 23 (2017). 
\bibitem{ref.redner_shrinking} P. L. Krapivsky and S. Redner, Am. J. Phys. {\bf 72} 591 (2004).
\bibitem{ref.Hassan_Dongen} P. G. J. van Dongen and M. H. Ernst,  Phys. Rev. Lett. {\bf 54} 1396 (1985)
\bibitem{ref.family}  F.  Family, T. Vicsek,  J. Phys. A: Math and Gen {\bf 18}  75 (1985). 
 \bibitem{ref.viscek} T. Vicsek, F. Family, Phys. Rev. Lett. {\bf 52}  1669 (1984).
\bibitem{ref.march} P. G. J.  van Dongen, M. H. Ernst,  Phys. Rev. Lett. {\bf 54} 1396 (1985). 
\bibitem{ref.hassan_ba_dc} S. M. K. Hassan, M Z. Hassan and N. I Pavel,
J. Phys. A: Math. Gen. {\bf 44}  175101 (2011).
\bibitem{ref.hassan_debashish} D. Sarker, L. Islam and M. K. Hassan,  Chaos, Solitons \& Fractals 
{\bf 132} 109591 (2020). 
\bibitem{ref.reichl} L. E. Reichl, {\it A Modern Course in Statistical Physics} (Wiley-Interscience Publication, USA, 1998).
\bibitem{ref.weiss} G.H. Weiss, {\it  Aspects and Applications of the Random Walk}, 
(North-Holland, Amsterdam, 1994).
\bibitem{ref.alexander_1} J. C. Alexander and J. A. Yorke,  Ergod. Theory Dyn. Syst. {\bf 4} 1 (1984)
\bibitem{ref.alexander_2} J. C. Alexander and D. Zagier,  J. Lond. Math. Soc. {\bf 44} 121 (1991).
\bibitem{ref.torre_shrinking} A. C. de la Torre, A. Maltz, H. O. Ma´rtin, P. Catuogno, and I. 
Garcı \'{a}-Mata, Phys. Rev. E {\bf 62} 7748 (2000). 

\end{thebibliography}
\end{document}